\documentclass[epj]{svjour}
\usepackage{graphics}
\usepackage[utf8]{inputenc}  
\usepackage{graphicx,color} 
\usepackage{mathrsfs}
\usepackage{bm}
\usepackage{amsmath,amssymb,amsfonts}
\usepackage{amssymb}
\usepackage{ulem}
\usepackage{dsfont}
\usepackage{adjustbox}
\usepackage{tikz}
\usetikzlibrary{calc,trees,positioning,arrows,chains,shapes.geometric,%
    decorations.pathreplacing,decorations.pathmorphing,shapes,%
    matrix,shapes.symbols}
\usetikzlibrary{arrows,positioning} 

\tikzset{
>=stealth',
  punktchain/.style={
    rectangle, 
    rounded corners, 
    draw=black, very thick,
    text width=9em, 
    minimum height=3em, 
    text centered, 
    on chain},
  line/.style={draw, thick, <->},
  element/.style={
    tape,
    top color=white,
    bottom color=blue!50!black!60!,
    minimum width=8em,
    draw=blue!40!black!90, very thick,
    text width=10em, 
    minimum height=3.5em, 
    text centered, 
    on chain},
  every join/.style={<->, thick,shorten >=1pt},
  decoration={brace},
  tuborg/.style={decorate},
  tubnode/.style={midway, right=2pt},
}

\tikzset{
     >=stealth',
     punkt/.style={
            rectangle,
            rounded corners,
            draw=black, very thick,
            text width=6.5em,
            minimum height=2em,
            text centered},
     pil/.style={
            ->,
            thick,
            shorten <=2pt,
            shorten >=2pt,}
 }

\newcommand{\sumint}{\int\hspace{-14pt}\sum}

\begin{document}
\title{Asymmetry dependence of Gogny based optical potential\thanks{Contribution to the Topical Issue "Finite range effective interactions 
and associated many-body methods - A tribute to Daniel Gogny" edited by Nicolas Alamanos, Marc Dupuis, Nathalie Pillet.}}
\author{G. Blanchon\inst{1,}\thanks{e-mail: guillaume.blanchon@cea.fr} \and M. Dupuis\inst{1} \and R. N. Bernard\inst{1} \and H. F. Arellano \inst{2,1} 
 %
}                     
%
%
\institute{CEA,DAM,DIF F-91297 Arpajon, France \and Department of Physics - FCFM, University of Chile, Av. Blanco Encalada 2008, Santiago, Chile}
\date{Received: date / Revised version: date}
%
\abstract{An analysis of neutron and proton scattering off $^{40,48}$Ca has been carried out. Real and imaginary potentials 
have been generated using the Nuclear Structure Method for scattering with the Gogny D1S nucleon-nucleon effective interaction. 
Observables are well described by NSM for neutron and proton elastic scattering off $^{40}$Ca and for neutron scattering off $^{48}$Ca.
For proton scattering off $^{48}$Ca, NSM yields a lack of absorption. This discrepancy is attributed to double-charge-exchange contribution 
and coupling to Gamow-Teller mode which are not included in the present version of NSM. A recipe based on a Perey-Buck fit of NSM imaginary 
potential and Lane model is proposed to overcome this issue in an approximate way.
\PACS{
      {PACS-key}{describing text of that key}   \and
      {PACS-key}{describing text of that key}
     } 
} 
\maketitle
\section{Introduction}

Optical potentials are key for the description of nucleon-nucleus direct elastic and inelastic scatterings \cite{satchler_83}. 
Moreover, they are used to generate transmission coefficients for statistical model of compound nucleus such as in 
Hauser-Feshbach formalism \cite{hauser_52} and beyond \cite{moldauer_61,engelbrecht_73}. Furthermore, knowledge of the isospin 
asymmetry dependence of the potential is convenient when calculating quasielastic charge-exchange processes. For evaluation 
purposes, optical potential is often fitted in order to reproduce a consistent set of reaction observables. Whenever experimental 
data are not available, one can ideally rely on more microscopic approaches such as Nuclear Field Theory \cite{bes_76,potel_13}, 
Energy Density Functionals (EDF) \cite{vinhmau_70,mizuyama_12c,blanchon_15,hao_15}, \textit{ab initio} approaches 
\cite{hagen_12,holt_16} or mixed approaches such as g-matrix effective interaction folded with EDF density \cite{dupuis_06,arellano_11}. 
Moreover microscopic approaches can yield a physical guidance for new parametrizations of phenomenological potentials 
providing form factors for volume and surface parts of the potential, energy dependence, nonlocality shape and parameters, 
dependence on the isospin asymmetry of the target nucleus. 

In previous attempts \cite{blanchon_15,blanchon_15b}, Nuclear Structure Method (NSM) \cite{vinhmau_70} has been successfully 
applied to describe nucleon scattering off $^{40}$Ca using Gogny D1S interaction as sole input. This method is based on Green's 
function formalism. In its current version, it allows the description of nucleon scattering off doubly-closed shell spherical 
nuclei including integral and differential cross sections as well as spin observables below about 30~MeV incident energy. An 
extension to target nuclei with pairing and deformed target nuclei will be possible in a near future thanks to EDF's extended 
reach. It is worth noting that Gogny interaction has been originally fitted on structure observables. The nice agreement with 
reaction observables provided by NSM is mainly due to the correct description of the target-nucleus radius and the good 
description of collective states of the target nucleus provided by the Random-Phase Approximation (RPA) for doubly-closed 
shell nuclei. 

In this work, we apply NSM to nucleon scattering off both $^{40}$Ca and $^{48}$Ca in order to study optical potential 
dependence on the isospin asymmetry of the target-nucleus. In Sec.~\ref{sec:formalism}, we present a brief reminder of the NSM 
formalism. In Sec.~\ref{sec:cross}, in order to assess the validity of NSM potential, we confront NSM reaction observables  
with data. In Sec.~\ref{sec:volint}, volume integrals for NSM imaginary potential are presented with a special focus on the 
isospin asymmetry of the target-nucleus. In Sec.~\ref{sec:pb}, general trends of NSM potential are extracted by fitting a 
Perey-Buck like equivalent potential \cite{perey_62}. Then an approximation based on Lane consistency \cite{lane_62} is used 
in order to recover the missing absorption in the case of proton scattering off $^{48}$Ca. Finally, Sec.~\ref{sec:conclusions} 
lists the conclusions of this study.

\section{Nuclear Structure Method}
\label{sec:formalism}
NSM formalism is presented in detail in Ref.~\cite{blanchon_15,blanchon_15b}. We briefly introduce here the key points of 
the formalism. Equations are presented omitting spin for simplicity. NSM potential, $V$, consists of two components,
\begin{equation}
 V = V^{HF} + \Delta V. \label{eq:v}
\end{equation}
The former is a mean-field potential; the latter is a polarization potential built from target-nucleus excitations. The 
explicit coupling of the elastic channel to those inelastic channels results in a loss of flux which is reflected in the 
imaginary part of the complex $\Delta V$ potential.\\
The HF potential in coordinate space reads
\begin{eqnarray}
V^{HF}(\textbf{r},\textbf{r'})= \int d\textbf{r}_{1} v(\textbf{r},\textbf{r}_{1}) \rho(\textbf{r}_{1})\delta(\textbf{r}-\textbf{r'})
-v(\textbf{r},\textbf{r'})\rho(\textbf{r},\textbf{r'}), \nonumber \\
\label{eq:vhf}
\end{eqnarray}
where $v$ is the effective NN interaction. $\rho(\textbf{r})$ and $\rho(\textbf{r},\textbf{r'})$ are the usual local and 
nonlocal densities \cite{blanchon_15}, respectively. Rearrangement contributions stemming from the density-dependent term 
of the interaction are also accounted for \cite{ring_04}. \\

The polarization potential, $\Delta V$ in Eq.~\eqref{eq:v}, is built coupling the elastic channel to the intermediate excited 
states of the target nucleus. Those excited states are described within the RPA formalism \cite{ring_04}. Both excited states 
and couplings are generated using the same effective NN interaction. The resulting potential is nonlocal, energy dependent and 
complex. Going into more details, the polarization contribution to the potential reads
\begin{equation}
 \Delta V = V^{PP}+V^{RPA}-2V^{(2)}, \label{eq:dv}
\end{equation}
where $V^{PP}$ and $V^{RPA}$ are contributions from particle-particle and particle-hole correlations, respectively. 
The uncorrelated particle-hole contribution, $V^{(2)}$, is accounted for once in $V^{PP}$ and twice in $V^{RPA}$. It is 
subtracted twice in order to avoid double counting. When using Gogny interaction, part of particle-particle correlations 
is already contained at the HF level. We use the same prescription as in Ref.~\cite{bernard_79}, omitting the real part 
of $V^{PP}$ while approximating the imaginary part of $V^{PP}$ by $\textrm{Im}\left[V^{(2)}\right]$. The resulting 
potential is not fully dispersive anymore because of this latter term which does not have any real counterpart. Then 
Eq.~\eqref{eq:dv} reduces to
\begin{equation}
 \Delta V = \textrm{Im}\left[V^{(2)}\right]+V^{RPA}-2V^{(2)}. \label{eq:sig-approx}
\end{equation}	
\noindent For nucleons with incident energy $E$, the RPA potential reads,
\begin{eqnarray}
  V^{RPA}({\bf r,r'},E)&=& \sum_{N\neq 0} \sumint_{\lambda}
  \bigg[{{n_{\lambda}}\over{E-\varepsilon_{\lambda}+E_{N}-i\Gamma(E_{N})}} \nonumber\\
  &+&{{1-n_{\lambda}}\over{E-\varepsilon_{\lambda}-E_{N}+i\Gamma(E_{N})}}\bigg]\nonumber\\
  &\times& \Omega^{N}_{\lambda}(\textbf{r})\Omega^{N}_{\lambda}(\textbf{r'}), \label{eq:vrpa}
\end{eqnarray}
where $n_{i}$ and $\varepsilon_{i}$ are occupation number and energy of the single-particle state $\phi_{i}$ in the HF field, 
respectively. The label $\lambda$ refers to the single-particle state of the intermediate particle \cite{blanchon_15b}. $E_{N}$ 
and $\Gamma(E_{N})$ represent the energy and the width of the $N^{th}$ excited state of the target, respectively. Additionally, 
\begin{eqnarray}
 \Omega^{N}_{\lambda}(\textbf{r}) = \sum_{(p,h)} \left[X^{N,(p,h)}F_{ph\lambda}(\textbf{r}) + Y^{N,(p,h)}F_{hp\lambda}(\textbf{r})\right],
\end{eqnarray}
where $X$ and $Y$ denote the usual RPA amplitudes and
\begin{equation}
 F_{ij\lambda}(\textbf{r}) = \int d^{3} \textbf{r}_{1} \phi^{*}_{i}(\textbf{r}_{1})v(\textbf{r},\textbf{r}_{1})\left[1-\textrm{\^P}\right]\phi_{\lambda}(\textbf{r})\phi_{j}(\textbf{r}_{1}),
 \label{eq:fij}
\end{equation}
where $\textrm{\^P}$ is a particle-exchange operator and $v$ is the same effective NN interaction as in Eq.~\eqref{eq:vhf}. 
The uncorrelated particle-hole contribution reads
\begin{eqnarray}
  V^{(2)}({\bf r,r'},E) &=& \frac{1}{2}\sum_{ij} \sumint_{\lambda} \bigg[{{n_{i}(1-n_{j})n_{\lambda}}\over{E-\varepsilon_{\lambda}+E_{ij}-i\Gamma(E_{ij})}}\nonumber\\
  &+&{{n_{j}(1-n_{i})(1-n_{\lambda})}\over{E-\varepsilon_{\lambda}-E_{ij}+i\Gamma(E_{ij})}}\bigg]\nonumber\\
  &\times&  F_{ij\lambda}(\textbf{r})F^{*}_{ij\lambda}(\textbf{r}'), \label{eq:vph}
\end{eqnarray}
with $E_{ij} = \varepsilon_{i}-\varepsilon_{j}$, the uncorrelated particle-hole energy. \\

Calculations are performed according to the following scheme:
\begin{itemize}
 \item[-] $V^{HF}$ in Eq.~\eqref{eq:vhf} is converge using Gogny D1S interaction \cite{berger_91}. It is done in coordinate 
 space by diagonalization in a 15~fm box to ensure the correct asymptotic behavior of single-particle states.
 
 \item[-] $V^{HF}$ is used to generate the intermediate single-particle state labeled $\lambda$ in Eqs.~\eqref{eq:vrpa} and 
 \eqref{eq:vph}. Both discrete and continuum spectra of the intermediate single-particle state are accounted for.
  
 \item[-] Target excited states are obtained solving RPA equations in a harmonic oscillator basis, including fifteen major 
 shells \cite{blaizot_77} and using Gogny D1S interaction. We account for RPA excited states with spin up to $J=8$. 

 \item[-] $V^{RPA}$ and $V^{(2)}$ are obtained using Gogny D1S interaction in Eqs.~\eqref{eq:vrpa}.

 \item[-] The first zero-energy $J^{\pi}=1^{-}$ excited state obtained with RPA, containing the spurious translational mode, 
 is removed from the calculation in Eqs.~\eqref{eq:vrpa} and \eqref{eq:vph}.
 
 \item[-] Escape and damping widths are simulated assigning a single phenomenological width, $\Gamma(E_{N})$, to RPA states 
 and uncorrelated particle-hole excitations in Eqs.~\eqref{eq:vrpa} and \eqref{eq:vph}. $\Gamma(E_{N})$ takes the value of 
 2, 5, 15 and 50~MeV, for excitation energies of 20, 50, 100 and 200~MeV, respectively. For incident energies above about 
 10~MeV where compound elastic contribution is negligible, cross sections are not very sensitive to the value chosen for the 
 width \cite{blanchon_15b}. 

 \item[-] Double-charge exchange $(p,n,p)$ or $(n,p,n)$ is not accounted for, thus intermediate single-particle state 
 $\lambda$ is of the same kind than the incident and the outgoing particle. 
 
 \item[-] The integro-differential Schrödinger equation for scattering with the nonlocal potential, $V$ in Eq.~\eqref{eq:v}, 
 is solved following the matrix inversion method exposed in the documentation of the \textsc{DWBA} code \cite{raynal_98}.
\end{itemize}
The HF propagator is dressed only once. For this reason the scheme is self-consistent at the HF level and only consistent when 
considering polarization contributions. 

\section{Microscopic potential and reaction data}
\label{sec:cross}

\begin{figure*}
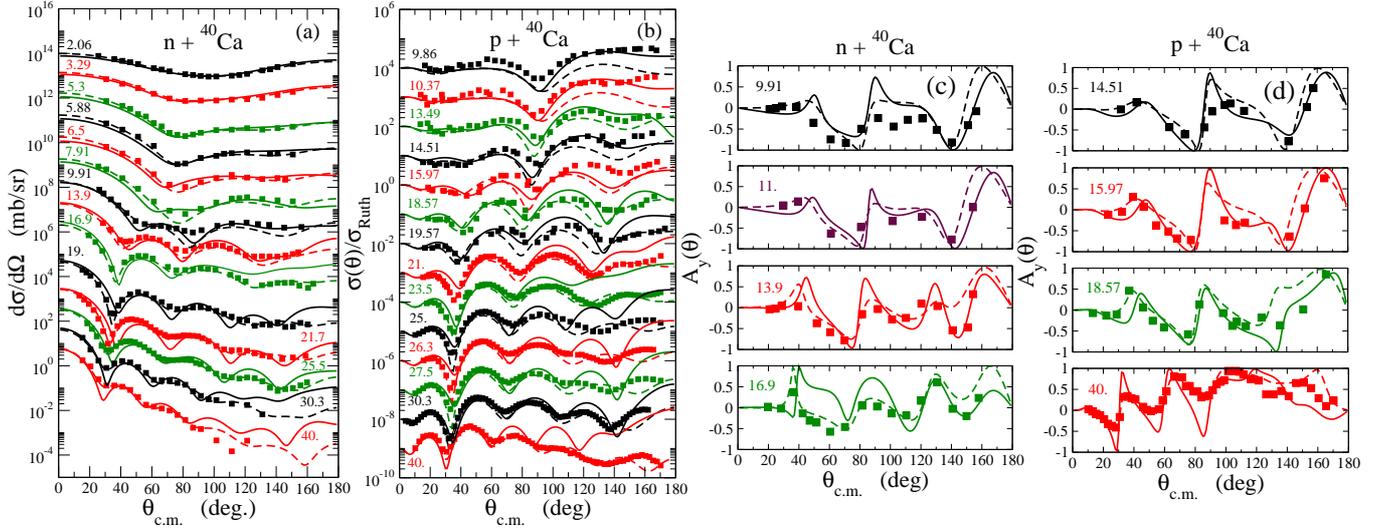

\begin{minipage}{0.24\textwidth}
\begin{center}
\adjustbox{trim={0.\width} {0.\height} {0.\width} {0.\height},clip}
{\includegraphics[width=1.05\textwidth,angle=-00,clip=false]{fig1.eps}}
\end{center}
\end{minipage}
\begin{minipage}{0.24\textwidth}
\begin{center}
\adjustbox{trim={0.\width} {0.\height} {0.\width} {0.\height},clip}
{\includegraphics[width=1.05\textwidth,angle=-00,clip=right]{fig2.eps}}
\end{center}
\end{minipage}
\begin{minipage}{0.24\textwidth}
\begin{center}
\adjustbox{trim={0.\width} {0.\height} {0.\width} {0.\height},clip}
{\includegraphics[width=1.05\textwidth,angle=-00,clip=false]{fig3.eps}}
\end{center}
\end{minipage}
\begin{minipage}{0.24\textwidth}
\begin{center}
\adjustbox{trim={0.\width} {0.\height} {0.\width} {0.\height},clip}
{\includegraphics[width=1.05\textwidth,angle=-00,clip=false]{fig4.eps}}
\end{center}
\end{minipage}
 \caption{Differential cross sections for neutron (a) and proton (b) scattering off $^{40}$Ca. Comparison 
 between data (symbols), $V^{HF}+\Delta V$ results (solid curves) and Koning-Delaroche potential results 
 (dashed curves). The same for analyzing powers for neutron (c) and proton (d) scattering. Incident energies 
 are indicated in MeV.}
 \label{fig:sec-40}
\end{figure*}

\begin{figure*}
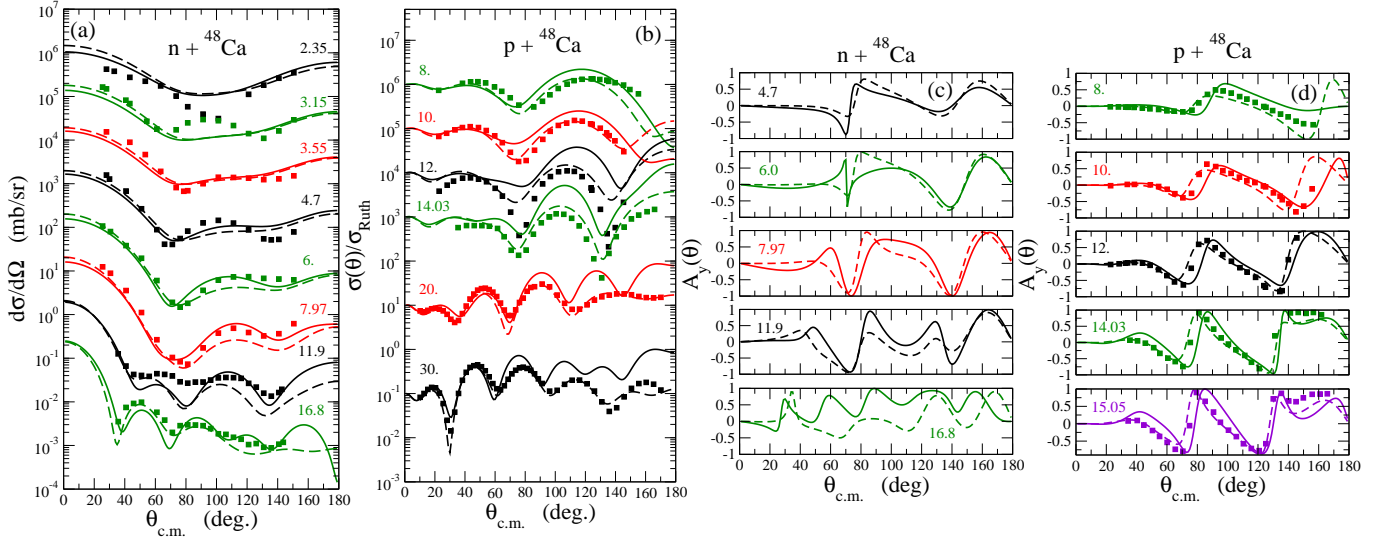

\begin{minipage}{0.24\textwidth}
\begin{center}
\adjustbox{trim={0.\width} {0.\height} {0.\width} {0.\height},clip}
{\includegraphics[width=1.05\textwidth,angle=-00,clip=false]{fig5.eps}}
\end{center}
\end{minipage}
\begin{minipage}{0.24\textwidth}
\begin{center}
\adjustbox{trim={0.\width} {0.\height} {0.\width} {0.\height},clip}
{\includegraphics[width=1.05\textwidth,angle=-00,clip=false]{fig6.eps}}
\end{center}
\end{minipage}
\begin{minipage}{0.24\textwidth}
\begin{center}
\adjustbox{trim={0.\width} {0.\height} {0.\width} {0.\height},clip}
{\includegraphics[width=1.05\textwidth,angle=-00,clip=false]{fig7.eps}}
\end{center}
\end{minipage}
\begin{minipage}{0.24\textwidth}
\adjustbox{trim={0.\width} {0.\height} {0.\width} {0.\height},clip}
{\includegraphics[width=1.05\textwidth,angle=-00,clip=false]{fig8.eps}}
\end{minipage} 
\caption{Same as Fig.~\ref{fig:sec-40} for $^{48}$Ca.}
 \label{fig:sec-48}
\end{figure*}

Before going any further into the analysis of NSM potential, we first check its ability to reproduce experimental reaction 
observables. NSM has been applied to neutron and proton scattering off $^{40-48}$Ca using Gogny D1S interaction. We focus on 
incident energies below 40~MeV where NSM has demonstrated to be efficient \cite{blanchon_15,blanchon_15b}. Results are 
summarized in Figs.~\ref{fig:sec-40} and \ref{fig:sec-48}. References to data are given in Ref.~\cite{koning_03}. 
Compound-elastic corrections furnished by the Hauser-Feshbach formalism \cite{hauser_52} using Koning-Delaroche potential 
\cite{koning_03} with \textsc{TALYS} \cite{koning_08} are applied to cross sections obtained from NSM and Koning-Delaroche 
potentials. It mainly implies a systematic correction on cross sections for neutron scattering with incident energy below 
10~MeV. 

\noindent In the case of neutron scattering off $^{40-48}$Ca (Figs.~\ref{fig:sec-40}a and \ref{fig:sec-48}a, respectively), 
NSM results compare very well to experiment and those based on Koning-Delaroche potential up to about 30~MeV incident energy. 
Regarding proton scattering, NSM yields nice results for $^{40}$Ca target up to 30~MeV incident proton energy 
(Fig.~\ref{fig:sec-40}b) but demonstrates an important lack of absorption at all incident energies in the case of $^{48}$Ca 
target (Fig.~\ref{fig:sec-48}b) even if the correct shape of the differential cross section is retained. 
\begin{figure}[h!]
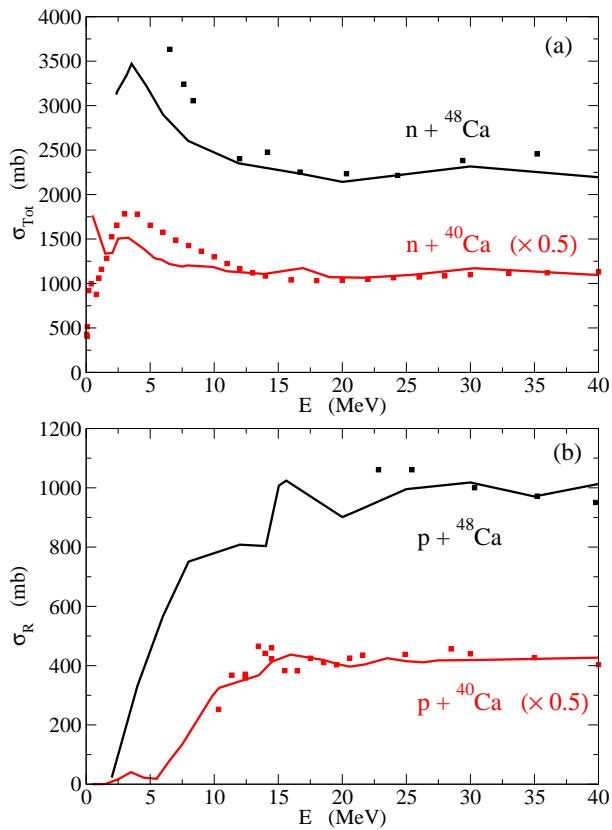

\begin{center}
\adjustbox{trim={0.\width} {0.\height} {0.\width} {0.\height},clip}
{\includegraphics[width=.9\linewidth,angle=-00,clip=false]{fig9.eps}}
\adjustbox{trim={0.\width} {0.\height} {0.\width} {0.\height},clip}
{\includegraphics[width=.9\linewidth,angle=-00,clip=false]{fig10.eps}}
\end{center}
\caption{Total cross section for neutron scattering off $^{40-48}$Ca (a). Reaction cross section for proton scattering off 
$^{40-48}$Ca (b).}
\label{fig:sec-tot}
\end{figure}
\noindent In Figs.~\ref{fig:sec-40}c and \ref{fig:sec-40}d, we show calculated analyzing powers for 
neutron and proton scattering off $^{40}$Ca, at several incident energies, in good agreement with experiment. In 
Fig.~\ref{fig:sec-48}d, analyzing powers for proton scattering off $^{48}$Ca compare quite well with experiment. 
Moreover, agreement with data is comparable to that obtained from Koning-Delaroche potential. These results suggest that NSM 
potential retains the correct spin-orbit behavior even in the case of proton-neutron asymmetry in the target-nucleus. 
To our knowledge, analyzing powers for neutron scattering off $^{48}$Ca are not available experimentally in this energy range. In 
Fig.~\ref{fig:sec-48}c, we present analyzing power predictions between 4.7~MeV and 16.8~MeV incident neutron energy. 
NSM description shows slight differences with Koning-Delaroche predictions.

\noindent In Fig.~\ref{fig:sec-tot}a, total cross sections for neutron scattering from both $^{40-48}$Ca targets
are in good agreement with experiment above about 10~MeV incident energy. Below that energy, total cross sections 
are underestimated by NSM. In Fig.~\ref{fig:sec-tot}b, we show reaction cross sections for proton scattering 
from both $^{40-48}$Ca targets. NSM results are in good agreement with experiment. In the case of $^{48}$Ca, 
only a few experimental data is available between 22~MeV and 40~MeV. \\
\begin{figure*}[t]
\adjustbox{trim={0.\width} {0.\height} {0.\width} {0.\height},clip}
{\includegraphics[angle=-00,width=1.0\textwidth]{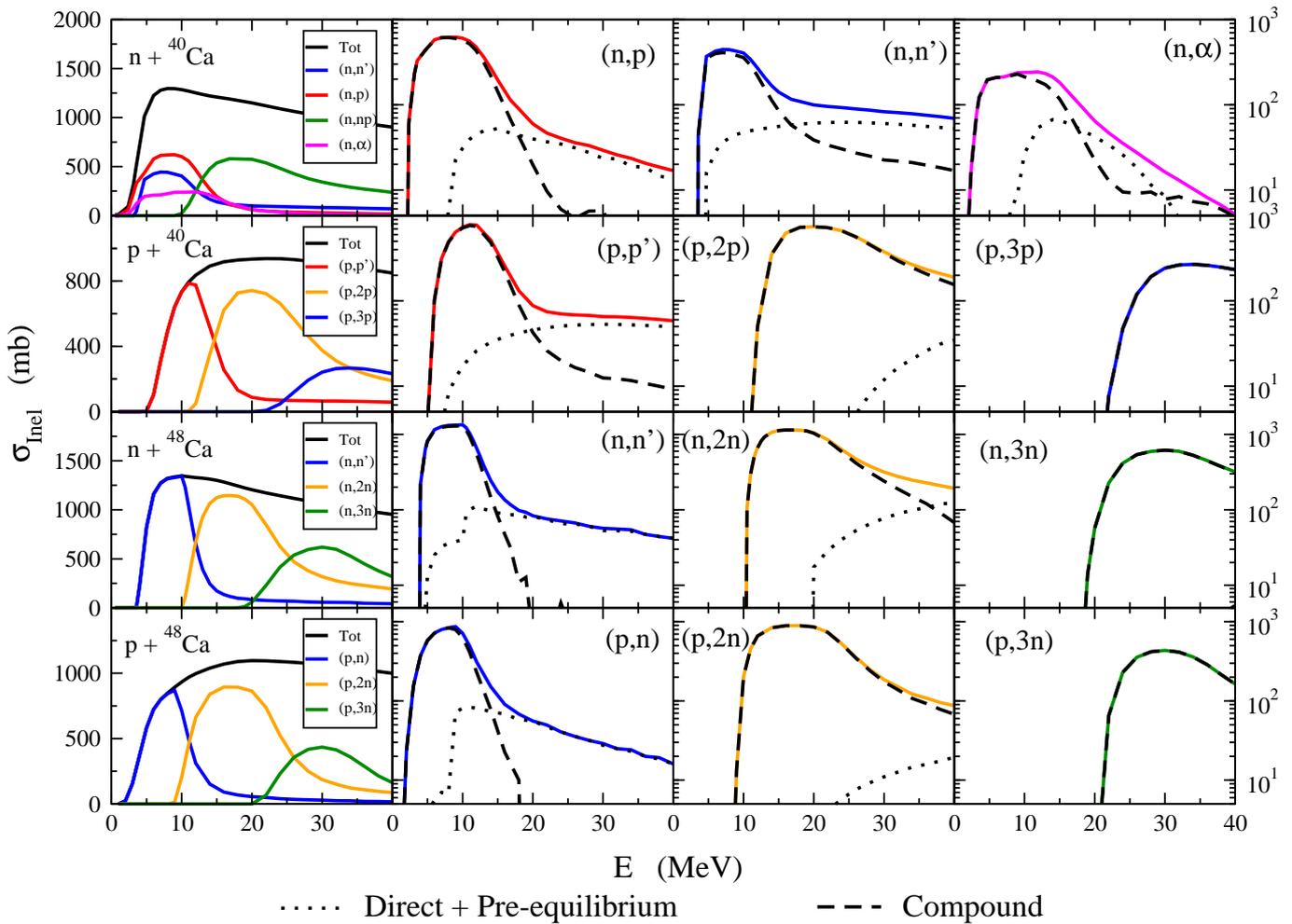}}
\caption{(Color online) \textsc{TALYS} evaluations of inelastic cross sections for neutron and proton scattering off 
$^{40-48}$Ca: Total reaction cross section (with linear scale on the left side) and the main components responsible for 
absorption (with logarithmic scale on the right side). Direct and Pre-equilibrium inelastic contributions are summed 
and are depicted in dotted line. Compound inelastic contribution are in dashed line.}
\label{fig:talys}
\end{figure*}

We wish now to investigate what makes NSM up to describe neutron and proton scattering off $^{40}$Ca 
and what makes it fail in reproducing proton scattering off $^{48}$Ca. 
\\

Charity \textit{et al}. have extensively explored 
the asymmetry dependence of dispersive optical potentials in Calcium isotopes \cite{charity_06,charity_07,charity_14}. They 
reach the conclusion that for proton elastic scattering, an increase in absorption is expected in going from $^{40}$Ca to 
$^{48}$Ca because of the coupling to the Gamow-Teller collective mode \cite{charity_07}. It is argued that this could enhance 
surface absorption with strength increasing as $\approx 3(N-Z)$. On the other hand, no change is expected for neutrons, as they 
do not couple to this resonance. In a further attempt, Waldecker \textit{et al}. \cite{waldecker_11} studied asymmetry from 
an \textit{ab-initio} point of view using Faddeev Random-Phase Approximation (FRPA) \cite{barbieri_01}. Results suggest that 
charge-exchange excitations of the target interfere only very weakly with the nucleon-nucleus scattering process. The most 
striking effect on the potential related to isospin asymmetry is related to the tensor term of the bare interaction which 
plays a major role on absorption in FRPA. 
\\

\noindent \textit{What does NSM contain?} In NSM, the $A+1$ system made of an incident nucleon and the $A$ nucleons of the 
target is described as a single-particle state in the HF field labeled $\lambda$ in Eq.~\eqref{eq:vrpa} and a RPA state 
describing target excitations at the one particle - one hole level. This description of the $A+1$ system allows to span over 
only part of the $A+1$ nucleon Hilbert space. As a result, at this level of approximation, NSM deals with processes involving 
0, 1 and 2 nucleons in the continuum, depending on whether the single-particle state is bound and the RPA excitation involves 
bound particle states, or the single-particle state is in the continuum or the RPA excitation involves continuum particle 
states, or both are in the continuum, respectively. NSM thus gives access to part of the direct emission of proton, neutron, 
neutron-proton pair and deuteron. In its present version, NSM does not explicitly account for double-charge exchange, meaning that the 
$\lambda$ intermediate single-particle state has the same isospin projection than the incident particle. This inhibits coupling 
to the Gamow-Teller collective mode as well as proton (neutron) pair direct emission in the case of neutron (proton) scattering. 
Moreover heavier composite particles such as tritium and $\alpha$ particles or more generally the direct emission of more than 
two nucleons would require a description of the target nucleus with higher order in perturbation such as second RPA 
\cite{gambacurta_10} or multiparticle-multihole configuration mixing \cite{pillet_07}. A direct reaction is a doorway leading 
toward compound nucleus formation. The compound nucleus is formed by a sequence of collisions, namely the pre-equilibrium, 
leading to increasingly complicated rearrangements of the target nucleus. NSM describes the first step of this sequence taking 
into account $2p-1h$ excitations with coherent particle-hole amplitudes provided by RPA. The phenomenological width applied 
to RPA states makes possible the damping toward compound nucleus. 
\\

\noindent \textit{NSM vs. evaluation.} Baring those considerations in mind, we study neutron and proton scattering off 
$^{40-48}$Ca targets below 40~MeV with the nuclear reaction code \textsc{TALYS} \cite{koning_08}. This evaluation tool deals 
with numerous reaction channels including direct, pre-equilibrium and compound reaction mechanisms. In particular, it provides 
a quantitative picture of the various particles emitted during the scattering process. The direct contribution is obtained 
through the coupling to experimentally known collective states. The corresponding coupled-channel problem is solved with ECIS 
code \cite{raynal_04}. The pre-equilibrium is obtained with the exciton model where target excitations are described within a 
particle-hole scheme. For incident energies below about 40~MeV, after primary pre-equilibrium emission the excitation energy of 
the residual nucleus is relatively small and one can safely assume that further decay of the nucleus proceeds mainly by compound 
mechanism \cite{koning_04}. The multiple pre-equilibrium emission is very weak. In Fig.~\ref{fig:talys}, we present the corresponding 
inelastic cross sections together with the main contributions to absorption. The distinction between the direct part and the 
pre-equilibrium mechanisms is rather arbitrary, especially under 20~MeV, so we present the sum of the two. We refer to it as direct 
contribution in the following.\\ 
For $^{40-48}$Ca targets within the energy range considered, \textsc{TALYS} results suggest that deuteron direct emission is 
negligible. As a result explicit coupling to intermediate deuteron is not mandatory in this study. \\
In the case of neutron scattering off $^{48}$Ca, the main components of the absorption ordered by crescent threshold 
energy are $(n,n')$, $(n,2n)$ and $(n,3n)$. The $(n,n')$ component is led by compound emission below about 15~MeV and by direct 
emission above. The following neutron multi-emission components are mostly of compound nature. Good results obtained 
with NSM are explained by the fact that $(n,n')$ is taken into account explicitly and it is the main doorway state. 
\\
The same logic holds for proton scattering off $^{40}$Ca where the main components of absorption are $(p,p')$, 
$(p,2p)$ and $(p,3p)$. The direct $(p,p')$ contribution is described within NSM and acts as a doorway that feeds the compound 
$(p,2p)$ and $(p,3p)$ contributions.
\\
For neutron scattering off $^{40}$Ca, the main components are $(n,p)$, $(n,n')$ and $(n,np)$. The $(n,np)$ 
contribution, not depicted in Fig.~\ref{fig:talys}, is mainly compound. The important contribution from $(n,p)$ is due to 
a low threshold energy for this reaction channel, about 500~keV, compared to the $(n,n')$'s one of about 3.5~MeV. Those last 
two components are partly contained in NSM where $(n,n')$ is explicitly taken into account and $(n,p)$ is partly accounted for 
with proton RPA excitations in the continuum. The inclusion of double charge exchange in the formalism would improve NSM 
prediction in that case. Moreover \textsc{TALYS} yields at low energy a $(n,\alpha)$ component with some direct contribution. 
The coupling to the $\alpha$ intermediate channel is beyond NSM's reach. Still, results obtained with NSM for neutron 
scattering off $^{40}$Ca are reasonable. 
\\
In the case of proton scattering off $^{48}$Ca, the absorption is mainly built from $(p,n)$, $(p,2n)$ and $(p,3n)$ 
channels. The last two contributions are mainly compound. $(p,n)$ channel has a direct component which is expected to be 
described only partially by NSM with neutron excitations in the continuum described with RPA. A great part of $(p,n)$ channel 
may be as well yielded by double-charge exchange. Once again this contribution is not accounted for in NSM which could explain 
the lack of absorption observed in differential elastic cross section for proton scattering off $^{48}$Ca in 
Fig.~\ref{fig:sec-48}b.
\\
To conclude this part, there are strong indications that coupling to Gamow-Teller mode would improve NSM description of 
proton scattering off $^{48}$Ca. More generally, coupling to Gamow-Teller mode should play an important role when dealing with 
proton scattering off neutron rich targets close to the neutron drip-line as neutron emission is favored.\\

\noindent \textit{NSM \& tensor interaction.} Waldecker \textit{et al.} pointed out the major influence of the tensor 
component of the bare interaction on the isospin asymmetry of the optical potential \cite{waldecker_11}. In this work, we use 
Gogny D1S interaction which doesn't contain any tensor component. Gogny interaction can be considered as a parametrization of a 
g-matrix. Part of the bare tensor contribution is included in particular in the central term of Gogny interaction \cite{bernard_16}. 
Moreover, elastic scattering calculations based on g-matrix have shown small differences with or without the g-matrix tensor 
component. Nonetheless, the addition of a tensor contribution in Gogny interaction has an effect both on the description of excited 
states with RPA \cite{anguiano_16} and on the coupling vertices in NSM. This effect of the tensor interaction has been investigated 
by Robin \textit{et al.} \cite{robin_16} within a relativistic formalism. Further consistent inclusion of a tensor contribution in 
Gogny interaction could help disentangle this issue. This is far beyond the scope of this work. \\

\section{Volume integrals}
\label{sec:volint}
Volume integrals are useful means of comparison between local potentials as they are well constrained by scattering data. 
When considering nonlocal potentials volume integrals are well constrained only in a multipole range depending on incident 
energy. Nevertheless, they still provide interesting information. For instance in a previous work \cite{blanchon_15}, volume integral 
of the real part of NSM potential shown to be well suited for incident energies below about 30 MeV and too attractive for 
energies beyond this limit. Here we focus on the imaginary part of NSM potential which is nonlocal and energy dependent. When 
solving the integro-differential Schrödinger equation, it is convenient to use a multipole expansion of the nonlocal potential 
\begin{equation}
V(\textbf{r}\sigma,\textbf{r'}\sigma';E) = \sum_{ljm} {\cal Y}_{ljm}({\bf \hat r} \sigma) \nu_{lj}(r,r';E){\cal Y}_{ljm}^{\dagger}({\bf \hat r'} \sigma'), 
\label{eq:pw_exp}
\end{equation}
with
\begin{equation}
{\cal Y}_{ljm}~({\bf \hat r} \sigma)~\equiv~[Y_{l}({\bf \hat r}) \otimes \chi_{1/2}(\sigma)]_{jm},
\end{equation}
with $Y_{l}^{m_{l}}(\hat r)$ the spherical harmonic and $\chi_{1/2}^{m_{s}}(\sigma)$ the spin function. In the case of a spherical 
target-nucleus, the potential is diagonal in $(l,j)$, the multipole expansion of Schrödinger equation is decoupled and it can 
be solved independently for each $(l,j)$ with the corresponding $\nu_{lj}(r,r';E)$ potential. The volume integral of the 
imaginary part of the nonlocal potential for a given multipole $(l,j)$ is defined as
\begin{eqnarray}
 J^{lj}_{W}(E) &=& \frac{-4\pi}{A} \int dr\ r^{2} \int dr' r'^{2} \textrm{Im} [\nu_{lj}(r,r',E)],
 \label{intvol}
\end{eqnarray}
where $A$ is the nucleon number of the target. In comparison, a local potential has volume integral which is independent 
of the multipole.
\begin{figure*}[t!]
\begin{center}
\adjustbox{trim={0.\width} {0.\height} {0.\width} {0.\height},clip}
{\includegraphics[angle=-00,width=.77\textwidth]{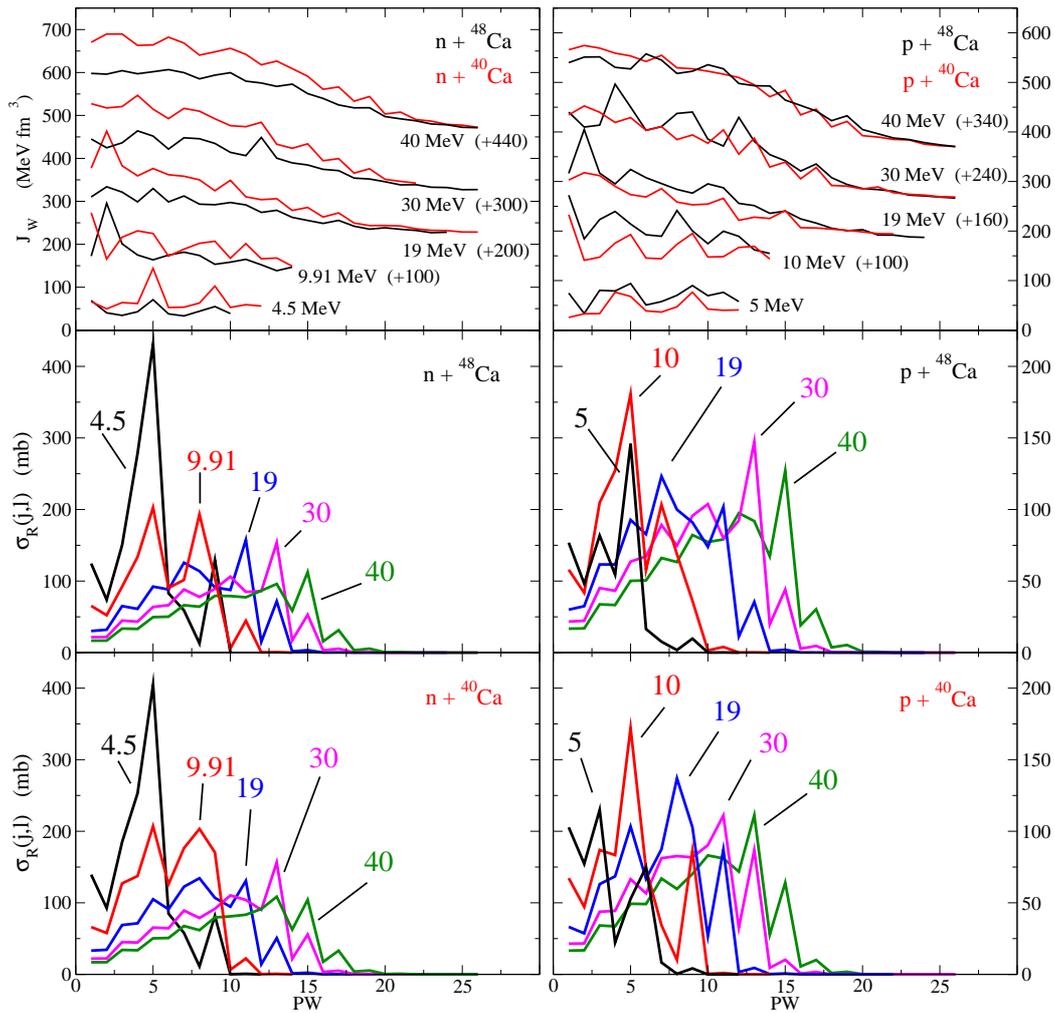}}
\caption{(color online) Volume integrals of the imaginary part of NSM potential for neutron and proton scattering off $^{40-48}$Ca (top panels) 
as function of partial waves. PW$=1,2,...$ stand for $(j,l)=(1/2,0)$, $(1/2,1)$, $(3/2,1)$..., partial waves, respectively. The lines have been offset 
along the $y$ axis by the indicated amounts. Reaction cross section as a function of multipole for the four reactions considered (four lower panels). 
Energies are indicated in MeV. Scales are different for each picture.}
\label{fig:intvol-40-48-np}
\end{center}
\end{figure*}
\noindent Volume integrals of the imaginary part of NSM potential for neutron and proton scattering off $^{40-48}$Ca are 
shown in top panels of Fig.~\ref{fig:intvol-40-48-np}. In the lower panel we present reaction cross sections for each 
multipole in order to emphasize multipoles contributing the most for each incident energy. At low incident energy the potential 
needs to be accurate for low multipoles as the Schrödinger equation is blind to what happens in higher multipoles and other 
energies. When incident energy increases the potential needs accuracy over higher multipole region. This partial-wave 
selectivity is the reason why an energy-dependent local potential, thus independent of the multipole, can be efficient in 
reproducing scattering observables. The only condition for a local potential to work is to be well tuned in the multipole region of 
interest for a given energy. \\
In the case of neutron scattering in Fig.~\ref{fig:intvol-40-48-np}, we observe a reduction of the volume integral 
going from $^{40}$Ca target to $^{48}$Ca using NSM. This effect is observed for incident neutrons at all energies considered 
below 40~MeV. This is a clear indication of the asymmetry of the neutron NSM potential. The asymmetry of the neutron potential 
in Calcium isotopes has been questioned by Charity \textit{et al.} fitting dispersive potential \cite{charity_07}. This study 
has motivated new experiments and a new optical potential analysis by Mueller \textit{et al.} \cite{mueller_11}. They study 
the surface and the volume magnitude of the potential and conclude that the neutron imaginary surface potential displays very 
little dependence on the neutron-proton asymmetry when going from $^{40}$Ca target to $^{48}$Ca one. Using radial integral 
rules for Woods-Saxon form factors, one shows that this behavior of the potential magnitude results in a depletion of about 
$6\%$ of the volume integral going from $^{40}$Ca to $^{48}$Ca. This is mainly due to the normalization by the total number 
of nucleons in Eq.~\eqref{intvol}. In Fig.~\ref{fig:intvol-40-48-np} the trend is the same in NSM and in the work of Mueller 
\textit{et al.} although asymmetry is more significant in NSM.\\
Below about 30~MeV incident energy, NSM potential for proton scattering provides the opposite trend than the one 
obtained with neutron with an enhancement of the volume integral going from $^{40}$Ca target to $^{48}$Ca. Increasing the 
incident energy, asymmetry vanishes and volume integrals for the two Calcium isotopes become comparable at about 40~MeV. Charity 
\textit{et al.} deduced the same behavior for proton scattering doing an optical model analysis of elastic scattering data 
\cite{charity_07}. This means that for proton scattering NSM already retains isospin asymmetry when including couplings to 
excited states of the target. The further inclusion of the double-charge exchange component shown to be important in proton 
scattering off $^{48}$Ca in Sec.~\ref{sec:cross} would allow for coupling to Gamow-Teller modes and should increase both absorption 
and volume integral.

\section{Perey-Buck equivalent potential}
\label{sec:pb}
\begin{figure*}
\begin{minipage}{0.24\linewidth}
\adjustbox{trim={0.\width} {0.\height} {0.\width} {0.\height},clip}
{\includegraphics[width=0.798\linewidth,angle=-90,clip=false]{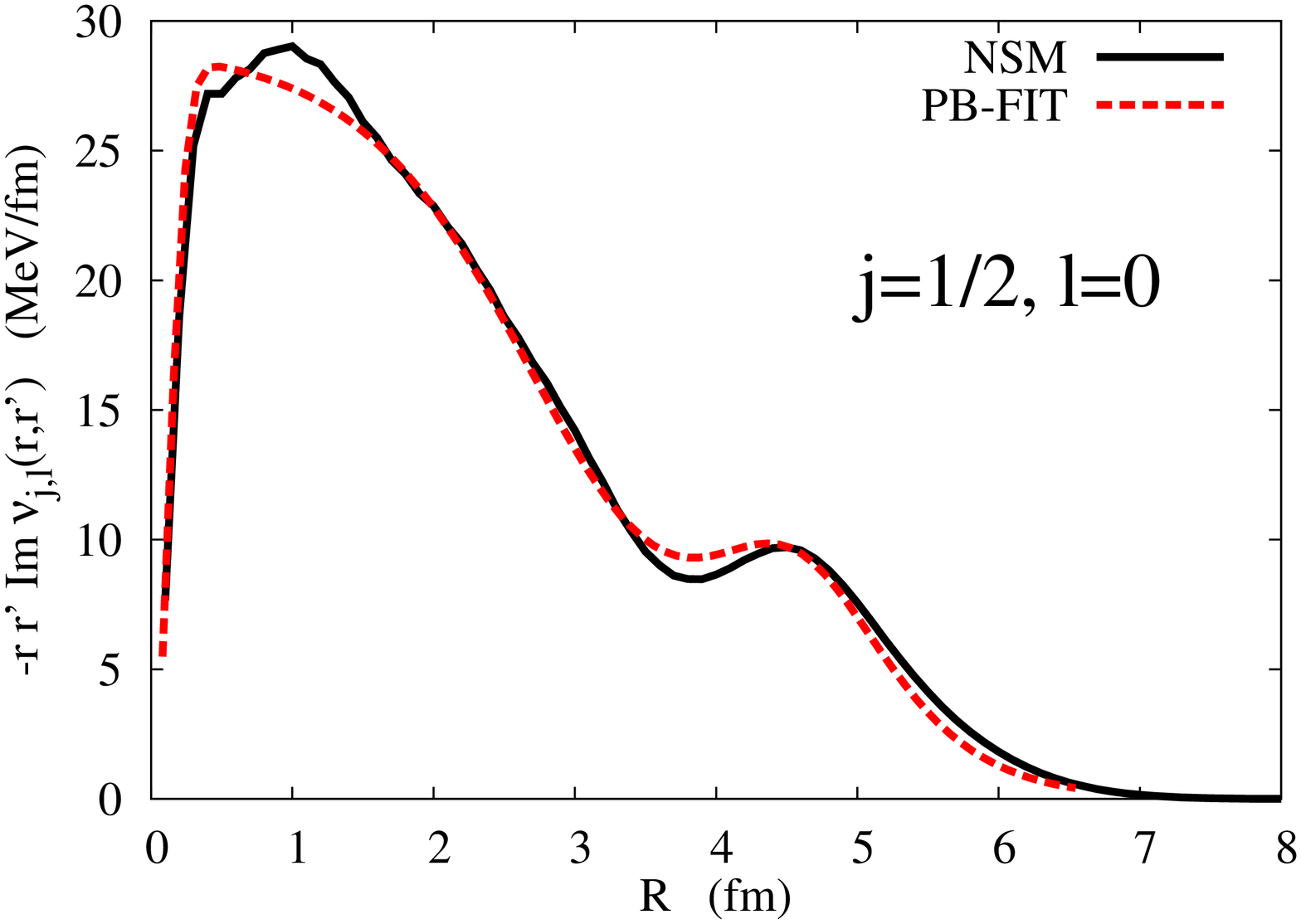}}\\
\adjustbox{trim={0.\width} {0.\height} {0.\width} {0.\height},clip}
{\includegraphics[angle=-90,width=1.145\linewidth,clip=false]{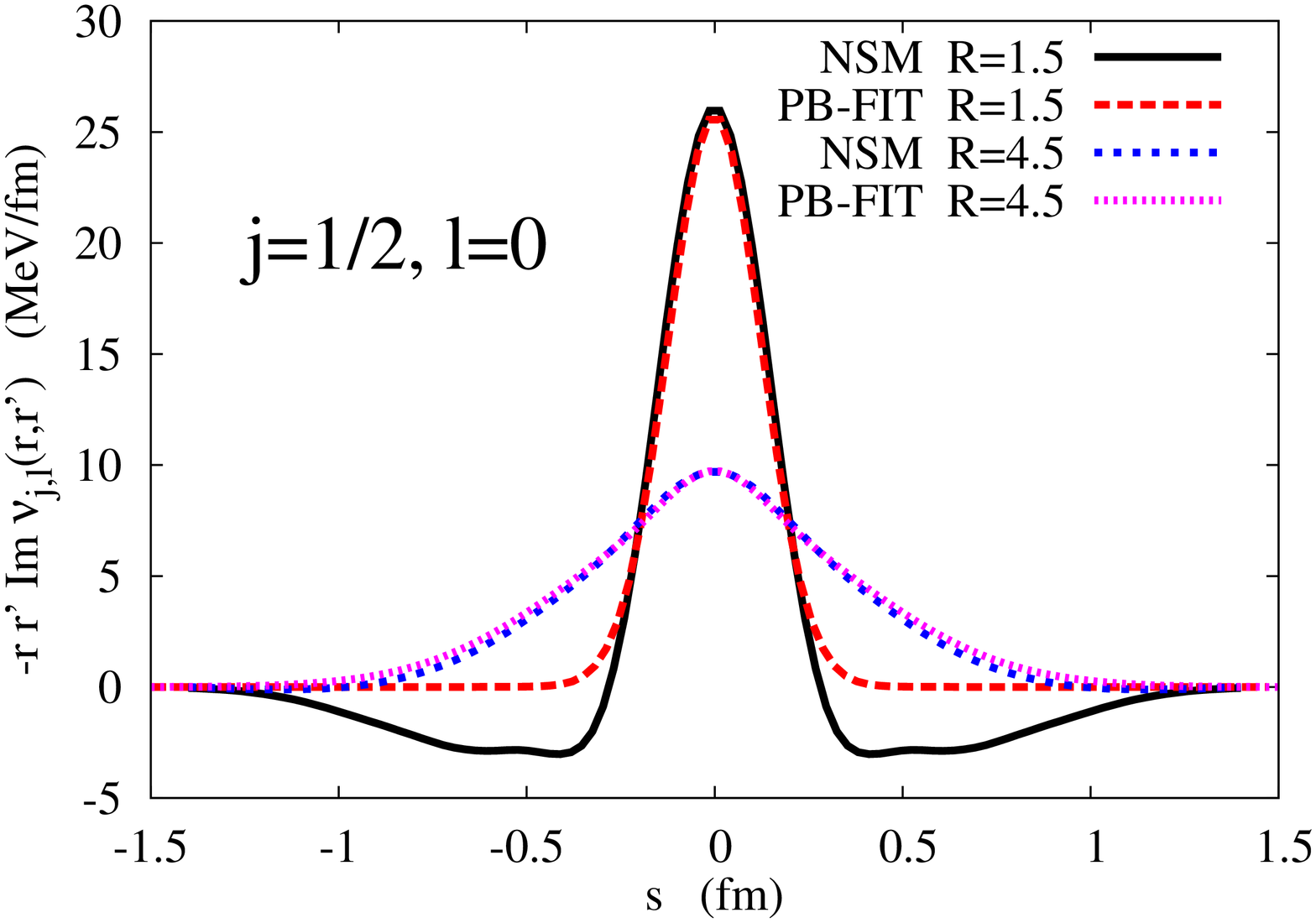}}
\end{minipage}
\begin{minipage}{0.24\linewidth}
\adjustbox{trim={0.\width} {0.\height} {0.\width} {0.\height},clip}
{\includegraphics[width=0.8\linewidth,angle=-90,clip=false]{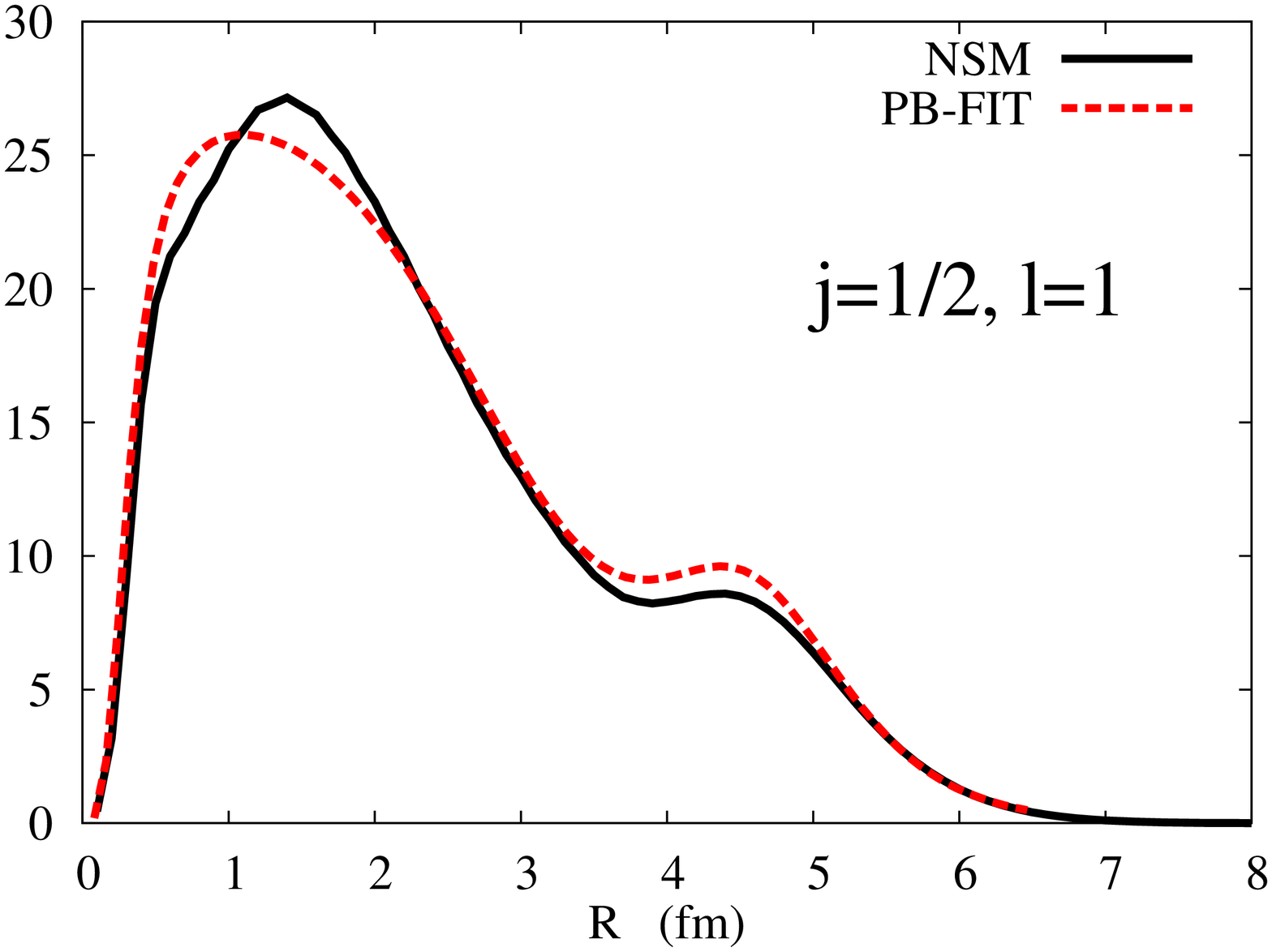}}\\
\adjustbox{trim={0.\width} {0.\height} {0.\width} {0.\height},clip}
{\includegraphics[angle=-90,width=1.15\linewidth,clip=false]{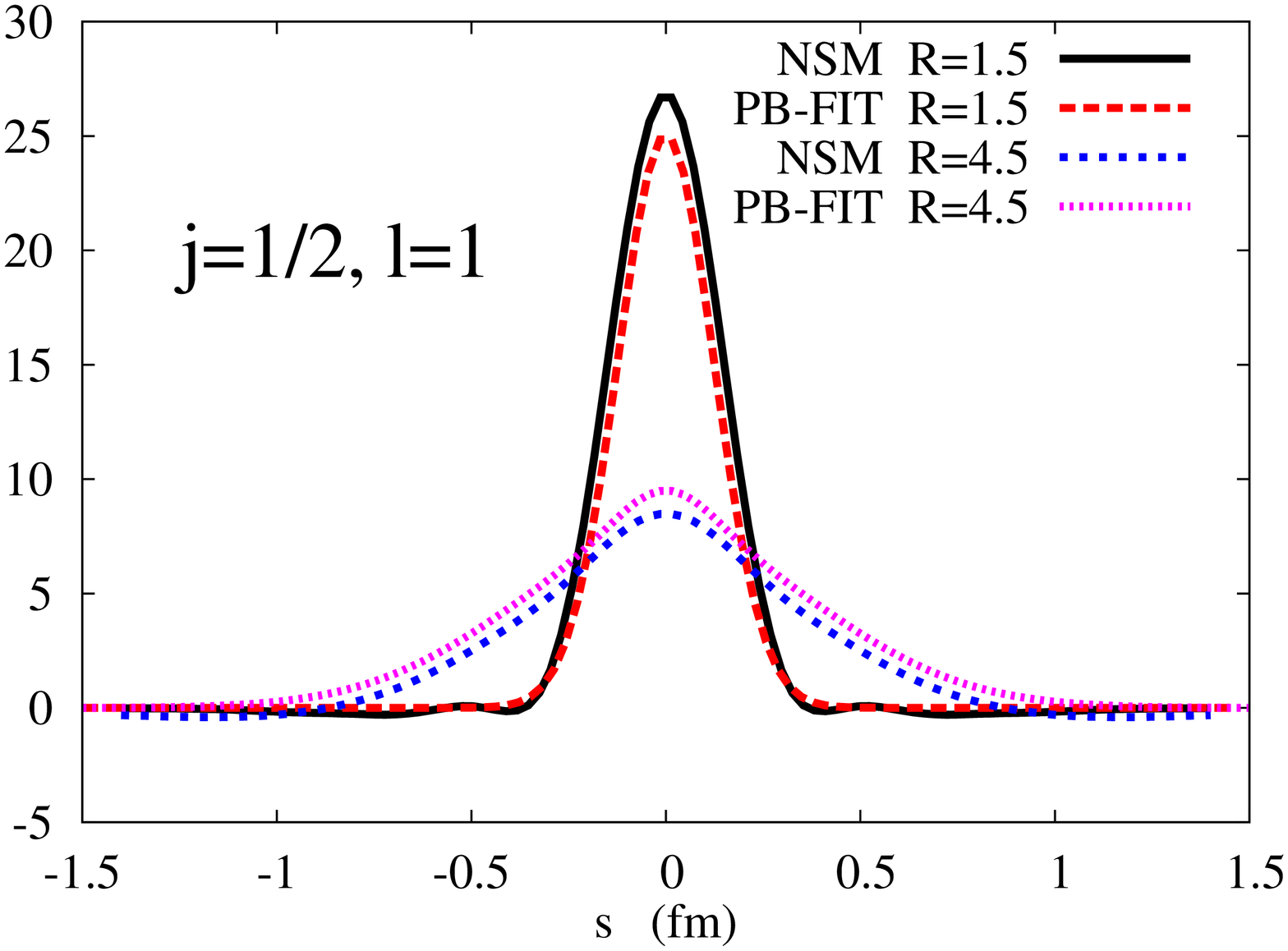}}
\end{minipage}
\begin{minipage}{0.24\linewidth}
\adjustbox{trim={0.\width} {0.\height} {0.\width} {0.\height},clip}
{\includegraphics[width=0.8\linewidth,angle=-90,clip=false]{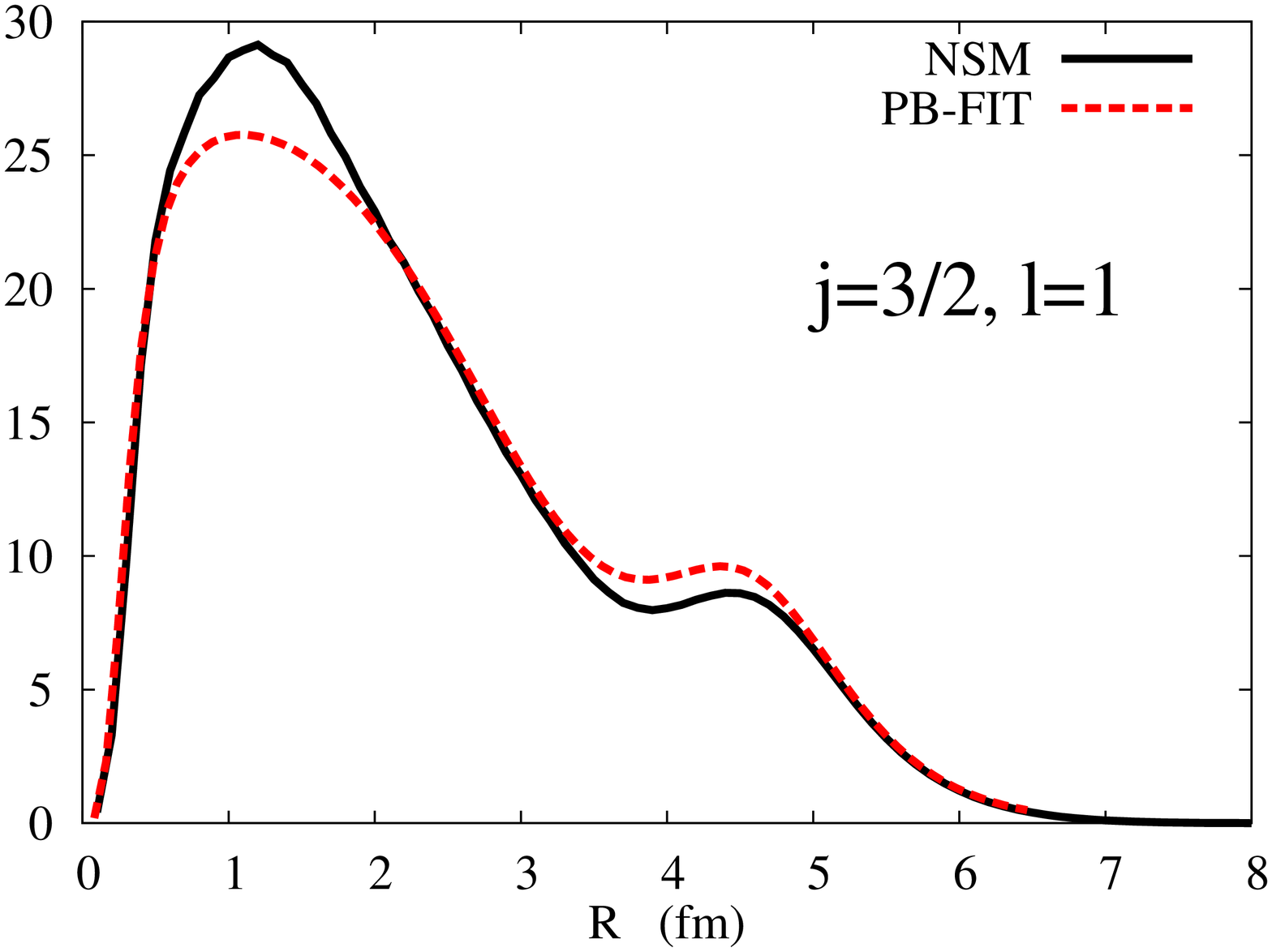}}\\
\adjustbox{trim={0.\width} {0.\height} {0.\width} {0.\height},clip}
{\includegraphics[angle=-90,width=1.15\linewidth,clip=false]{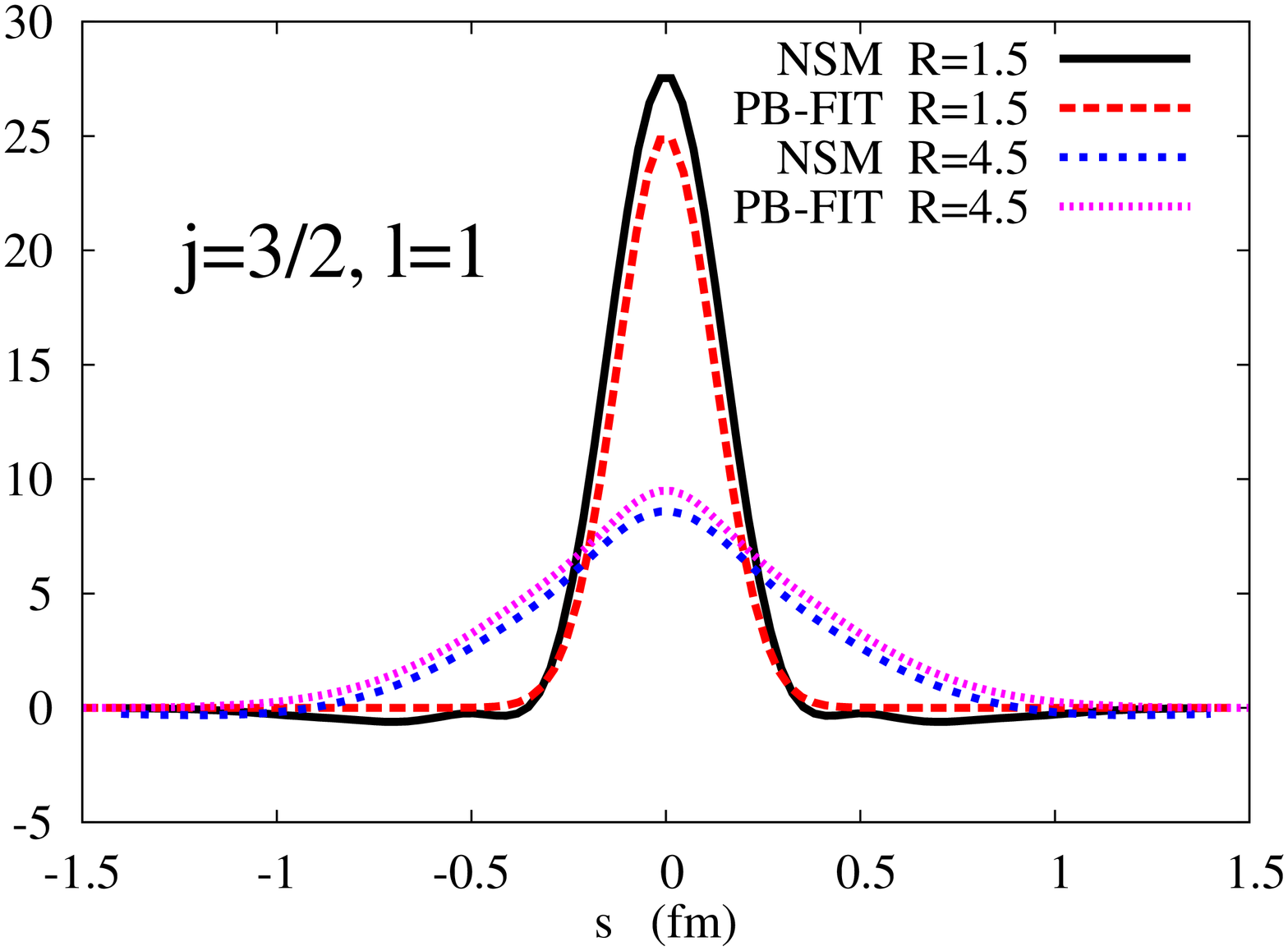}}
\end{minipage}
\begin{minipage}{0.24\linewidth}
\adjustbox{trim={0.\width} {0.\height} {0.\width} {0.\height},clip}
{\includegraphics[width=0.8\linewidth,angle=-90,clip=false]{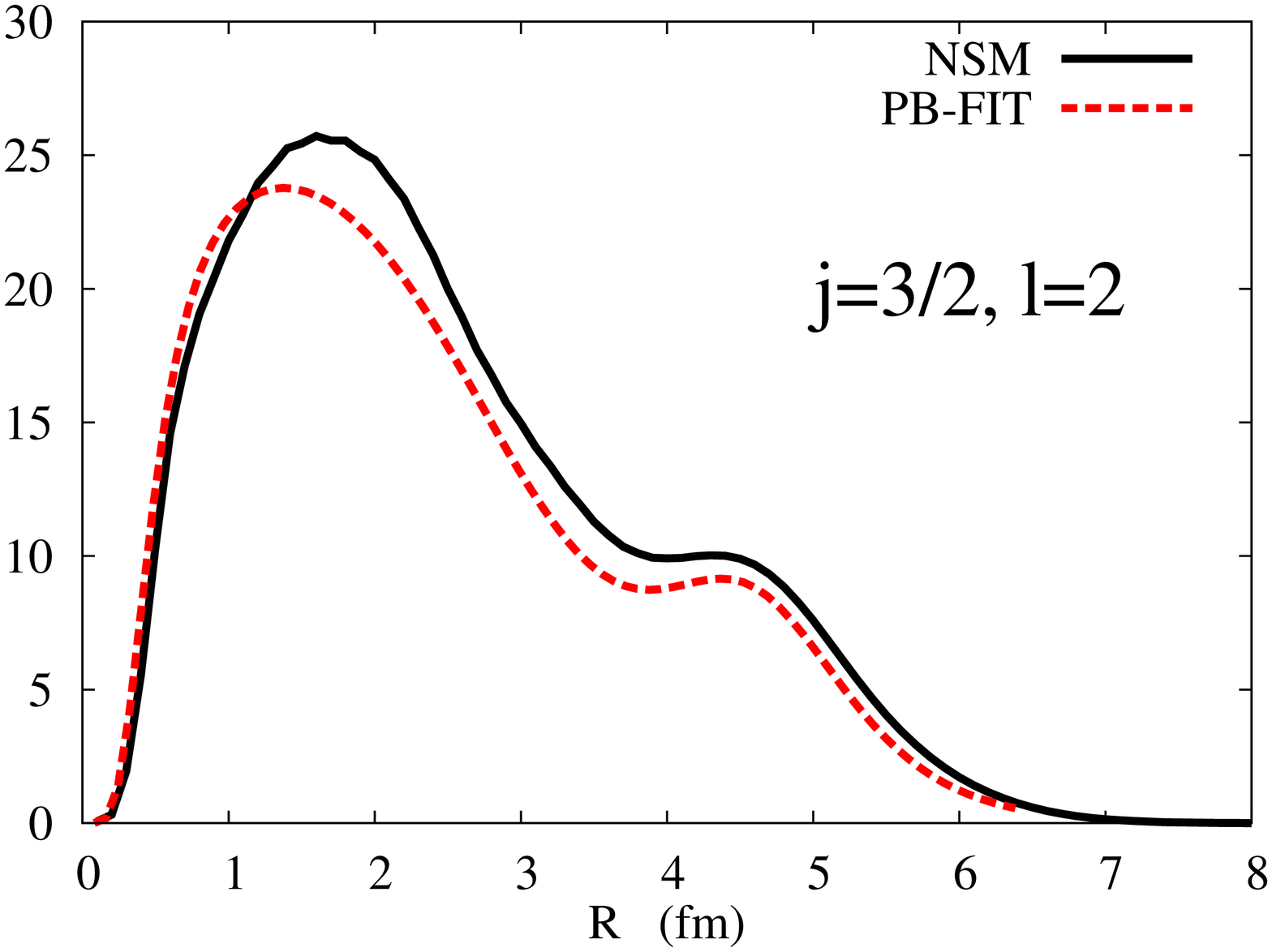}}\\
\adjustbox{trim={0.\width} {0.\height} {0.\width} {0.\height},clip}
{\includegraphics[angle=-90,width=1.15\linewidth,clip=false]{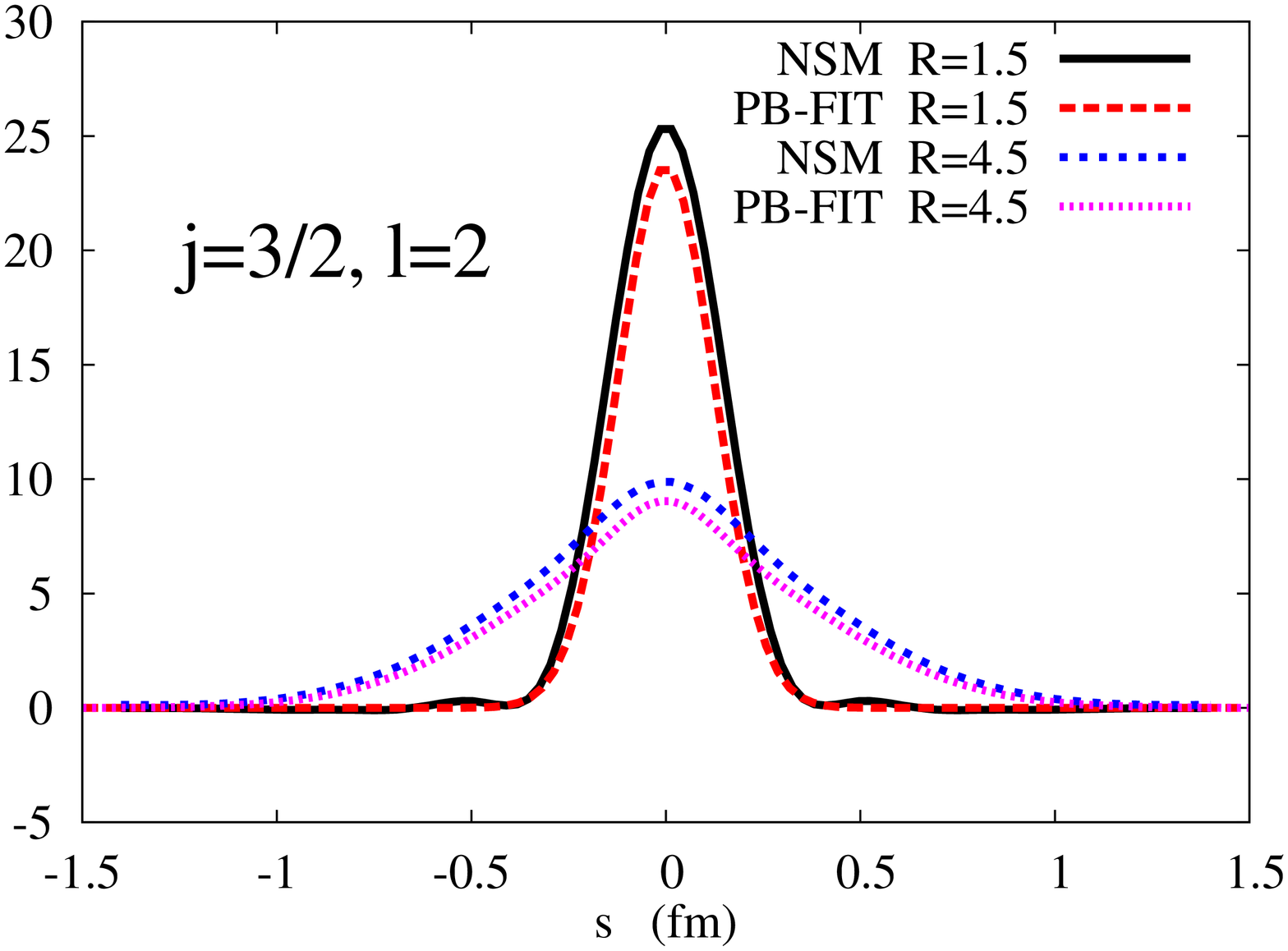}}
\end{minipage}
\caption{Sample of the PB-like potential fit of NSM for the first four multipoles for proton scattering off $^{48}$Ca at 30~MeV. 
(top panels) Diagonal contribution of the potential multipole expansion: NSM and fit. (bottom panels) Corresponding nonlocal 
contribution at $R=1.5$~fm and $R=4.5$~fm: NSM and fit.
}
\label{fig:fit}
\end{figure*}
\begin{figure}
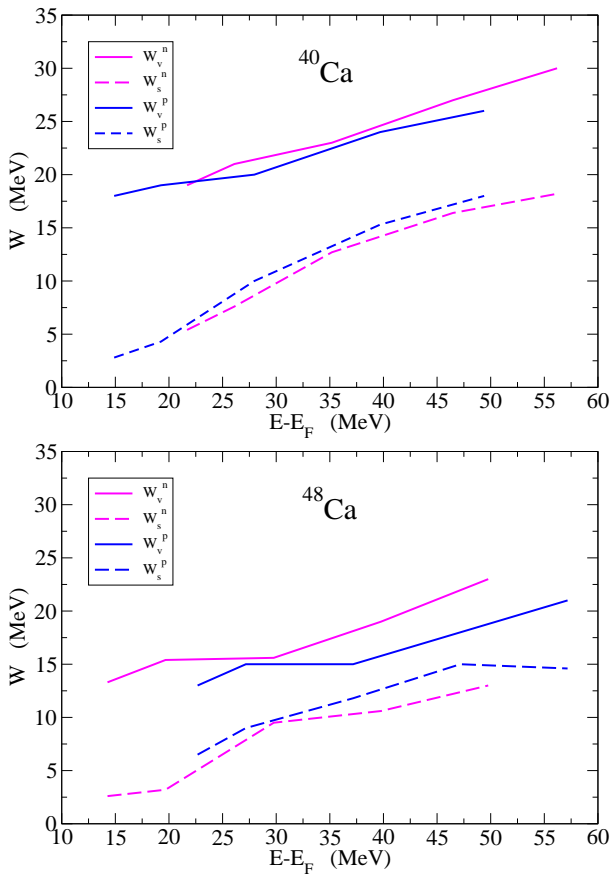

\begin{center}
\adjustbox{trim={0.\width} {0.\height} {0.\width} {0.\height},clip}
{\includegraphics[width=0.9\linewidth,angle=-00,clip=false]{fig22.eps}}\\
\adjustbox{trim={0.\width} {0.\height} {0.\width} {0.\height},clip}
{\includegraphics[width=0.9\linewidth,angle=-00,clip=false]{fig23.eps}}
\end{center}
\caption{Magnitude of surface part and volume part of Perey-Buck like potential fitted from NSM for neutron and proton
scattering off $^{40-48}$Ca.}
\label{fig:magn}
\end{figure}
\begin{figure}
\begin{center}
\adjustbox{trim={0.\width} {0.\height} {0.\width} {0.\height},clip}
{\includegraphics[width=0.9\linewidth,angle=-00,clip=false]{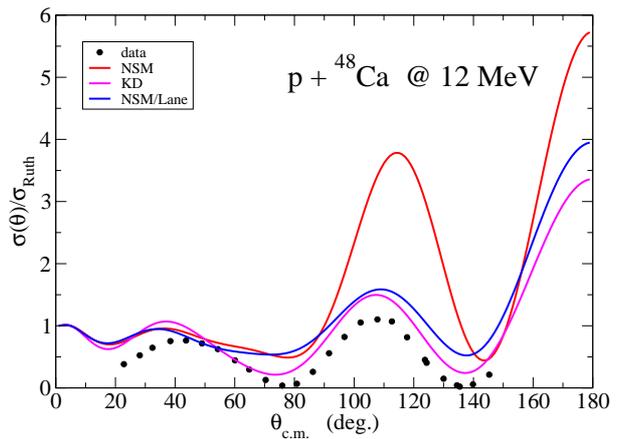}}\\
\end{center}
\caption{Differential elastic cross sections obtained with NSM (red line), Koning-Delaroche global potential (magenta line) and 
NSM with Lane correction (blue line) compared with data for proton scattering off $^{48}$Ca at 12~MeV.}
\label{fig:nsmlane}
\end{figure}

Presently, NSM ability to reproduce elastic scattering observables is encouraging but not competitive enough from the point 
of view of evaluation standards. Nevertheless, NSM can provide some guidance for further nonlocal potential parametrizations. 
Moreover there is a renewal of interest in the community for nonlocal potentials and their impact on reaction calculations 
\cite{mahzoon_14,titus_14,ross_15,titus_16,ross_16}. In this section, we present results of the fit of NSM imaginary 
potential using a Perey-Buck (PB) like form factor \cite{perey_62}. The potential ansatz is built from a volume part and 
a surface part, each one with a different Gaussian nonlocality. Hopefully the fitted potential shall yield cross sections 
equivalent to the ones obtained with NSM. In practice, PB-like imaginary potential reads
\begin{eqnarray}
 W^{n/p}(\textbf{r},\textbf{r'};E) &=& H(\textbf{s},\beta_{v}) W^{n/p}_{v}(E-E^{n/p}_{F}) f(R,r_{v},a_{v}) \nonumber \\ 
               &+& 4 a_{s} H(\textbf{s},\beta_{s}) W^{n/p}_{s}(E-E^{n/p}_{F})f'(R,r_{s},a_{s}), \nonumber \\
\end{eqnarray} 
where we use notations $R=(r+r')/2$ and $\textbf{s}=\textbf{r}-\textbf{r'}$. $v$ ($s$) subscript refers to parameters 
attributed to the volume (surface) part of the potential. $W^{n/p}_{v/s}(E-E^{n/p}_{F})$ is the energy-dependent magnitude 
of the potential, it varies for incident neutron and proton. $E^{n/p}_{F}$ is the Fermi energy. $\beta$ is the nonlocality 
parameter. Volume term is built from a Woods-Saxon form factor,
\begin{eqnarray}
f(r,r_{0},a) &=& \left[1+\exp\left(\frac{r-r_{0}A^{1/3}}{a}\right)\right]^{-1},
\end{eqnarray}
and surface term from a Woods-Saxon derivative with respect to $r$, $f'(r)$. $r_{0}$ and $a$ are 
reduced radius and diffuseness, respectively. The nonlocal form factor reads
\begin{equation}
 H(\textbf{s},\beta) = \frac{1}{\pi^{3/2}\beta^{3}} \exp\left(-\left|\frac{\textbf{s}}{\beta}\right|\right).
\end{equation}
We consider a Gaussian shape as in the original Perey-Buck phenomenological potential. Moreover this shape has already shown 
to provide a good description of NSM imaginary potential nonlocality \cite{blanchon_15b}. This is worth mentioning that the 
Gaussian shape obtained with NSM is not related to the use of Gogny interaction which is itself built from Gaussian form 
factors. The same nonlocality shape is obtained using for example Skyrme interactions \cite{bernard_79,bouyssy_81}. 
In the present work, two different nonlocality parameters are required in order to describe the surface and the volume 
contributions of the potential. Then following Perey~\textit{et al.} \cite{perey_62} prescription, making the approximation 
$|\textbf{r}+\textbf{r'}|\approx(r+r')$  allows for the multipole decomposition of the potential as shown in 
Eq.~\eqref{eq:pw_exp}
\begin{equation}
 H_{l}(r,r',\beta) = \frac{2i^{l}z}{\pi^{\frac{1}{2}}\beta}j_{l}(-iz)\exp\left(-\frac{r^{2}+r'^{2}}{\beta^{2}}\right),
\end{equation}
where $j_{l}$ are spherical Bessel functions and $z=2rr'/\beta^{2}$. Finally the multipole expansion of the potential reads
\begin{eqnarray}
 W_{l}^{n/p}(r,r';E) &=& H_{l}(r,r',\beta_{v}) W^{n/p}_{v}(E-E^{n/p}_{F}) f(R,r_{v},a_{v}) \nonumber \\ 
               &+& 4 a_{s} H_{l}(r,r',\beta_{s}) W^{n/p}_{s}(E-E^{n/p}_{F})f'(R,r_{s},a_{s}).\nonumber \\
\label{eq:fit}
\end{eqnarray}
In a first attempt we do not consider any imaginary spin-orbit contribution. As a consequence the multipole expansion is 
only $l$-dependent.
\begin{table}[h!]
\caption {PB-like parametrization of NSM for neutron and proton projectiles for $^{40-48}$Ca targets. All parameters are 
expressed in $fm$.}
\begin{center}
\begin{tabular}{ccccccc}
\hline
\hline
     $r_{v}$               &    $r_{s}$     &                  $a_{v}$                     &      $a_{s}$      &        $\beta_{v}$     &     $\beta_{s}$      \\
\hline
       0.78                &      1.254     &   0.49 ($^{40}$Ca)      0.78 ($^{48}$Ca)     &        0.44       &          0.35          &        1.1           \\
\hline
\hline
\end{tabular}
\end{center}
\label{table:fit}
\end{table}
Fitted parameters obtained for incident energies below 40~MeV are gathered in Table~\ref{table:fit}. Except for the diffuseness 
of the volume component, a single parameter set is obtained for both $^{40}$Ca and $^{48}$Ca. \\
We find a different reduced radius for the surface and the volume contribution. Surface reduced radius, $r_{s}$, is close 
to the value generally adopted in optical potential analyses. Volume reduced radius, $r_{v}$, is quite small. This small 
value of the volume radius is somehow compensated by a deeper volume magnitude. \\ 
Regarding nonlocalities NSM predicts a smaller nonlocality parameter for the volume part of the potential and a larger one 
for the surface. Mahzoon~\textit{et al.} found the same trend \cite{mahzoon_14}. \\
Despite the fact that PB-like potential is too a simple ansatz to represent NSM potential, it provides a rough 
idea of the shape of the potential. In Fig.~\ref{fig:fit} we present a sample of the fit for proton scattering off 
$^{48}$Ca at 30~MeV. For imaginary potential we adopt positive values when absorptive. For sake of concision we only present 
the first four multipoles of the potential but a reasonable agreement is obtained as well for higher multipoles. Diagonal 
($r=r'$) contributions are fitted using the ansatz of Eq.~\eqref{eq:fit}. They are presented in the top panels of 
Eq.~\eqref{fig:fit}. The surface and the volume nonlocality parameters are adjusted as well by looking at two transversal 
cuts of the nonlocal potential at $R=1.5$~fm and $R=4.5$~fm, respectively. Corresponding results are presented in the lower 
panels of Fig.~\ref{fig:fit}. In the first multipole some emissive "wings" appears in the nonlocality at different incident 
energies. We do not take this effect into account in order to keep the fit form factor as simple as possible. Moreover, the 
first multipole is expected to play a major role mainly at very low scattering energy. \\
\begin{figure}
\begin{center}
\adjustbox{trim={0.\width} {0.\height} {0.\width} {0.\height},clip}
{\includegraphics[width=0.9\linewidth,clip=false]{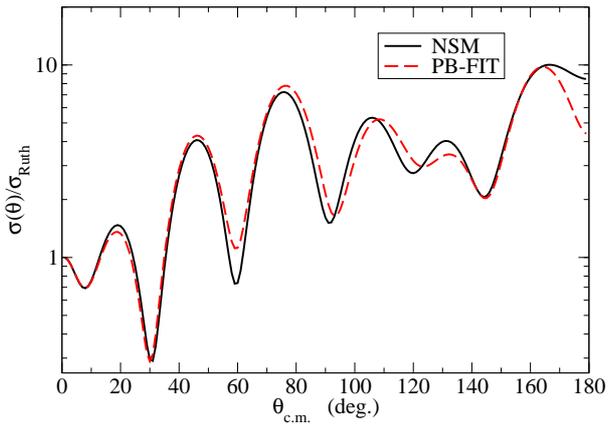}}
\end{center}
\caption{Comparison between differential elastic cross section for proton scattering off $^{48}$Ca at 30~MeV obtained with NSM 
and with the equivalent PB-like potential.}
\label{fig:fitsec}
\end{figure} 
Then we check the ability of the PB-like potential to describe differential elastic cross section obtained with NSM. 
As shown in Fig.\ref{fig:fitsec} in the case of proton scattering off $^{48}$Ca at 30~MeV, agreement is good enough to consider 
that the fit retains the main features of NSM potential. The same agreement is obtained for all the cases discussed. Then one can 
explore the behavior of the magnitudes for the surface and the volume contributions as a function of energy and projectile and look 
for evidences of asymmetry. Results of the fit of the magnitudes for neutron and proton scattering for $^{40-48}$Ca are summarized 
in Fig.~\ref{fig:magn}. Magnitudes are depicted as a function of incident energy subtracted by Fermi energy given in Table~\ref{table:efermi}. 
For simplicity we use Fermi energies yielded by the HF calculation which corresponds to the energy of the last fully occupied orbital. 
In reality one should determine the Fermi energy taking into account the energy dependent RPA contribution to the real part but at 
that stage the implementation is not yet suited in order to describe negative energies. Nevertheless, this contribution is expected 
to be small. \\
Results for $^{40}$Ca (Fig.~\ref{fig:magn} top panel) show that neutron and proton magnitudes both follow the same trend for 
surface and volume components. This is what is expected from a Lane consistent potential for scattering off self-conjugate 
target nucleus \cite{lane_62}. The same behavior has been observed by Mueller \textit{et al.} \cite{mueller_11} without imposing 
Lane consistency during the fit procedure. \\
Going from $^{40}$Ca to $^{48}$Ca the proton surface magnitude is increased whereas the neutron one follows the opposite trend. 
In the meantime, volume magnitudes for both neutron and proton are reduced of about 10~MeV. Once again this is a proof of 
asymmetry already retained  at the level of the coupling to excited states of the target. As we have seen in 
Sec.~\ref{sec:cross}, NSM leads to a lack of absorption in the description of proton scattering off $^{48}$Ca. By comparison 
with \textsc{TALYS}, it has been attributed to the absence of double-charge-exchange $(p,n,p)$ contribution avoiding coupling 
to Gamow-Teller mode. So we can expect asymmetry in the proton case to be enhanced when including double-charge exchange. 
Considering neutron magnitudes the observed asymmetry with NSM is in disagreement with results obtained by Mueller 
\textit{et al.} \cite{mueller_11} who predicted no asymmetry going from $^{40}$Ca to $^{48}$Ca for neutron scattering.\\
\begin{table}
\caption {HF Fermi energies in MeV.}
\begin{center}
\begin{tabular}{ccccccc}
\hline
\hline
     $E_{F}^{n}$     $\textit{(A=40)}$      &    $E_{F}^{p}$    $\textit{(A=40)}$ &    $E_{F}^{n}$      $\textit{(A=48)}$     &    $E_{F}^{p}$     $\textit{(A=48)}$    \\
\hline
                   -16.19                   &                -9.39                &                -9.77                      &                -17.17                   \\
\hline
\hline
\end{tabular}
\end{center}
\label{table:efermi}
\end{table}
In the Lane model \cite{lane_62}, which assumes isospin symmetry in nuclei, the nucleon-nucleus potential can be decomposed 
into isoscalar $V_{0}$ and isovector  $V_{1}$ parts,
\begin{equation}
V^{(n/p)} = V_{0} \pm \frac{(N-Z)}{2A} V_{1},
\label{eq:lane}
\end{equation}
where $(+)$ stands for neutron projectile and $(-)$ for proton projectile. $N$ ($P$) is the neutron (proton) number 
in the target nucleus. As already discussed by Osterfeld \textit{et al.} \cite{osterfeld_81b}, NSM contains 
by construction Lane inconsistent terms. Indeed the difference between neutron potential and proton potential, 
\begin{equation}
  V^{n}-V^{p} = \frac{N-Z}{A} V_{1} - V_{cc},
\label{eq:lanediff}
\end{equation}
can be due not only to the isospin conserving Lane potential as stated in Eq.~\eqref{eq:lane} in but also to isospin 
nonconserving Coulomb corrections, $V_{cc}$. These Coulomb corrections stem from the second-order part of the potential. Both 
theoretical \cite{jlm_77} and empirical \cite{rapaport_80} findings are in agreement that Coulomb correction weakens the 
absorption for protons compared to neutrons of the same energy. Nevertheless, Osterfeld \textit{et al.} have shown that Coulomb 
correction is small in the case of $^{40}$Ca \cite{osterfeld_81b} what is confirmed by our results on magnitude in 
Fig.~\ref{fig:magn}. Hence one can safely consider Eq.~\eqref{eq:lane} as a reasonable approximation for $^{40}$Ca in the 
considered energy range. On the opposite, Coulomb correction is not negligible in the case of $^{48}$Ca \cite{osterfeld_85}. 
Neglecting Coulomb correction in Eq.~\eqref{eq:lanediff}, one can in principle recover an upper limit of the absorption for 
proton scattering off $^{48}$Ca. \\
In the case of nucleon scattering off $^{40}$Ca ($N~=~Z~=~20$), Lane model leads to 
the same potential for neutron and proton projectile. One can parametrize the $^{40}$Ca potential using the ansatz of 
Eq.~\eqref{eq:fit}. Then using the same parametrization with $A=48$ one gets the isoscalar part of the potential $V^{FIT}_{0}$ 
extrapolated for $^{48}$Ca. Assuming NSM reasonably describes neutron scattering off $^{48}$Ca as shown in Sec.~\ref{sec:cross}, 
one finally gets NSM/Lane version of the proton-$^{48}$Ca potential,
\begin{equation}
 V^{p}_{NSM/Lane}(^{48}\mathrm{Ca}) = 2 V^{FIT}_{0}(^{48}\mathrm{Ca}) - V^{n}_{NSM}(^{48}\mathrm{Ca}),
 \label{eq:nsmlane}
\end{equation}
where NSM potential is used for the neutron-$^{48}$Ca potential. As a first test case in Fig.~\ref{fig:nsmlane} we present 
the differential elastic cross section for proton scattering off $^{48}$Ca at 20~MeV. NSM/Lane prescription is used for 
the imaginary part of the potential whereas the real part is kept as the one used in NSM calculations presented in 
Sec.~\ref{sec:cross}. The use of NSM/Lane potential described in Eq.~\eqref{eq:nsmlane} greatly improve the description 
of the cross section by substantially enhancing the absorption. Cross section presented in linear scale are very close from 
the Koning-Delaroche benchmark. \\ 
Higher-order versions of Lane prescription have been investigated by Holt~\textit{et al.} \cite{holt_16} and 
could be relevant in the $^{48}$Ca case. Nevertheless, one has to keep in mind that formally NSM potential contains 
Lane-inconsistent terms in the second order terms. So one should carefully take into account the double charge exchange 
in order to get a reliable description of the potential. \\

\section{Conclusions}
\label{sec:conclusions}
A study of neutron and proton scattering off $^{40}$Ca and $^{48}$Ca targets has been undertaken using Nuclear Structure 
Method for scattering. Gogny D1S nucleon-nucleon effective interaction is used consistently throughout the determination of 
the optical potential. NSM provides a reasonable description of neutron scattering off $^{40-48}$Ca and proton scattering off 
$^{40}$Ca below about 30~MeV incident energy. On the other hand a default of absorption is observed in differential cross 
sections for proton scattering off $^{48}$Ca for all considered incident energies between 8~MeV and 30~MeV. An independent 
calculation using \textsc{TALYS} evaluation tool has shown the importance of direct neutron emission in that case. This result 
points out the importance of double-charge exchange in proton scattering off $^{48}$Ca. This is most probably the case in 
general for proton scattering off neutron-rich targets inclined to emit neutrons. \\
The study of multipole dependent volume integrals of nonlocal NSM imaginary  contribution reveals that NSM 
predicts asymmetry for neutron scattering going from $^{40}$Ca target to the $^{48}$Ca one. The opposite trend is observed 
for proton scattering below about 30~MeV. Above this energy value the asymmetry tends to diminish leading to comparable 
at about 40~MeV. Results obtained for proton scattering should be modified when accounting for $(p,n,p)$ double-charge exchange. \\
Then we have proceeded to a fit of the NSM imaginary potential with a Perey-Buck like potential extracting in particular  
magnitudes for neutron and proton projectiles and for $^{40}$Ca and $^{48}$Ca targets. Results for neutron and proton 
potentials for scattering off $^{40}$Ca tend to demonstrate that Lane-inconsistent terms in NSM potential are small in that 
case. Based on that observation, we have used Lane prescription in order to successfully account for the lack of absorption 
observed in proton scattering off $^{48}$Ca. Nevertheless, NSM potential contains Lane-inconsistent terms by construction and 
the issue of double-charge exchange will have to be faced in further attempts following for example work from 
Osterfeld \textit{et al.} \cite{osterfeld_81b}. 

Study of heavier double-closed-shell target-nuclei is in progress. Moreover the extension to target nuclei experiencing 
pairing using Gorkov formalism has been initiated for spherical targets. This will give access to hundred of new targets.

\section*{Acknowledgments}
\noindent H. F. A. acknowledges partial funding from FONDECYT under Grant No 1120396.

\bibliographystyle{epj}

\begin{thebibliography}{47}

\bibitem{satchler_83}
G.~Satchler, \emph{Direct Nuclear Reactions}, International series of
  monographs on physics (Clarendon Press, 1983)

\bibitem{hauser_52}
W.~Hauser, H.~Feshbach, Phys. Rev. \textbf{87}, 366 (1952)

\bibitem{moldauer_61}
P.A. Moldauer, Phys. Rev. \textbf{123}, 968 (1961)

\bibitem{engelbrecht_73}
C.A. Engelbrecht, H.A. Weidenm\"uller, Phys. Rev. C \textbf{8}, 859 (1973)

\bibitem{bes_76}
D.~B\`es, R.~Broglia, G.~Dussel, R.~Liotta, R.~Perazzo, Nuclear Physics A
  \textbf{260}(1), 77  (1976)

\bibitem{potel_13}
G.~Potel, A.~Idini, F.~Barranco, E.~Vigezzi, R.A. Broglia, Reports on Progress
  in Physics \textbf{76}(10), 106301 (2013)

\bibitem{vinhmau_70}
N.~Vinh~Mau, Theory of nuclear structure (IAEA, Vienna) p. 931 (1970)

\bibitem{mizuyama_12c}
K.~Mizuyama, K.~Ogata, Phys. Rev. C \textbf{86}, 041603 (2012)

\bibitem{blanchon_15}
G.~Blanchon, M.~Dupuis, H.F. Arellano, N.~Vinh~Mau, Phys. Rev. C \textbf{91},
  014612 (2015)

\bibitem{hao_15}
T.V.N. Hao, B.M. Loc, N.H. Phuc, Phys. Rev. C \textbf{92}, 014605 (2015)

\bibitem{hagen_12}
G.~Hagen, N.~Michel, Phys. Rev. C \textbf{86}, 021602 (2012)

\bibitem{holt_16}
J.W. Holt, N.~Kaiser, G.A. Miller, Phys. Rev. C \textbf{93}, 064603 (2016)

\bibitem{dupuis_06}
M.~Dupuis, S.~Karataglidis, E.~Bauge, J.P. Delaroche, D.~Gogny, Phys. Rev. C
  \textbf{73}(1), 014605 (2006)

\bibitem{arellano_11}
H.F. Arellano, E.~Bauge, Phys. Rev. C \textbf{84}, 034606 (2011)

\bibitem{blanchon_15b}
G.~Blanchon, M.~Dupuis, H.F. Arellano, Eur. Phys. J. A \textbf{51}(12), 165
  (2015)

\bibitem{perey_62}
F.~Perey, B.~Buck, Nucl. Phys. \textbf{32}, 353  (1962)

\bibitem{lane_62}
A.~Lane, Nuclear Physics \textbf{35}, 676  (1962)

\bibitem{ring_04}
P.~Ring, P.~Schuck, \emph{The Nuclear Many-Body Problem}, Physics and astronomy
  online library (Springer, 2004)

\bibitem{bernard_79}
V.~Bernard, N.~Van~Giai, Nucl. Phys. \textbf{A327}(2), 397  (1979)

\bibitem{berger_91}
J.F. Berger, M.~Girod, D.~Gogny, Comput. Phys. Commun. \textbf{63}, 365 (1991)

\bibitem{blaizot_77}
J.~Blaizot, D.~Gogny, Nucl. Phys. \textbf{A284}(3), 429  (1977)

\bibitem{raynal_98}
J.~Raynal, computer code DWBA98, 1998, (NEA 1209/05)

\bibitem{koning_03}
A.J. Koning, J.P. Delaroche, Nucl. Phys. \textbf{A713}(3-4), 231  (2003)

\bibitem{koning_08}
A.J. Koning, S.~Hilaire, M.~Duijvestijn, in \emph{Proceeding of the
  International Conference on Nuclear Data for Science and Technology-ND2007}
  (EDP Sciences, Paris, France, 2008), pp. 211--214

\bibitem{charity_06}
R.J. Charity, L.G. Sobotka, W.H. Dickhoff, Phys. Rev. Lett. \textbf{97}, 162503
  (2006)

\bibitem{charity_07}
R.J. Charity, J.M. Mueller, L.G. Sobotka, W.H. Dickhoff, Phys. Rev. C
  \textbf{76}(4), 044314 (2007)

\bibitem{charity_14}
R.J. Charity, W.H. Dickhoff, L.G. Sobotka, S.J. Waldecker, The European
  Physical Journal A \textbf{50}(2), 1 (2014)

\bibitem{waldecker_11}
S.J. Waldecker, C.~Barbieri, W.H. Dickhoff, Phys. Rev. C \textbf{84}, 034616
  (2011)

\bibitem{barbieri_01}
C.~Barbieri, W.H. Dickhoff, Phys. Rev. C \textbf{63}, 034313 (2001)

\bibitem{gambacurta_10}
D.~Gambacurta, M.~Grasso, F.~Catara, Phys. Rev. C \textbf{81}, 054312 (2010)

\bibitem{pillet_07}
N.~Pillet, N.~Sandulescu, P.~Schuck, Phys. Rev. C \textbf{76}, 024310 (2007)

\bibitem{raynal_04}
J.~Raynal, computer code ECIS03, 2004 (NEA 0850/16)

\bibitem{koning_04}
A.~Koning, M.~Duijvestijn, Nuclear Physics A \textbf{744}, 15  (2004)

\bibitem{bernard_16}
R.N. Bernard, M.~Anguiano, Nuclear Physics A \textbf{953}, 32  (2016), ISSN
  0375-9474

\bibitem{anguiano_16}
M.~Anguiano, A.M. Lallena, G.~Co', V.~De~Donno, M.~Grasso, R.N. Bernard, The
  European Physical Journal A \textbf{52}(7), 1 (2016)

\bibitem{robin_16}
C.~Robin, E.~Litvinova, The European Physical Journal A \textbf{52}(7), 205
  (2016)

\bibitem{mueller_11}
J.M. Mueller, R.J. Charity, R.~Shane, L.G. Sobotka, S.J. Waldecker, W.H.
  Dickhoff, A.S. Crowell, J.H. Esterline, B.~Fallin, C.R. Howell et~al., Phys.
  Rev. C \textbf{83}, 064605 (2011)

\bibitem{mahzoon_14}
M.H. Mahzoon, R.J. Charity, W.H. Dickhoff, H.~Dussan, S.J. Waldecker, Phys.
  Rev. Lett. \textbf{112}, 162503 (2014)

\bibitem{titus_14}
L.J. Titus, F.M. Nunes, Phys. Rev. C \textbf{89}, 034609 (2014)

\bibitem{ross_15}
A.~Ross, L.J. Titus, F.M. Nunes, M.H. Mahzoon, W.H. Dickhoff, R.J. Charity,
  Phys. Rev. C \textbf{92}, 044607 (2015)

\bibitem{titus_16}
L.J. Titus, F.M. Nunes, G.~Potel, Phys. Rev. C \textbf{93}, 014604 (2016)

\bibitem{ross_16}
A.~Ross, L.J. Titus, F.M. Nunes, Phys. Rev. C \textbf{94}, 014607 (2016)

\bibitem{bouyssy_81}
A.~Bouyssy, H.~Ng{\^o}, N.~Vinh~Mau, Nucl. Phys. \textbf{A371}(2), 173  (1981)

\bibitem{osterfeld_81b}
F.~Osterfeld, V.A. Madsen, Phys. Rev. C \textbf{24}, 2468 (1981)

\bibitem{jlm_77}
J.P. Jeukenne, A.~Lejeune, C.~Mahaux, Phys. Rev. C \textbf{16}(1), 80 (1977)

\bibitem{rapaport_80}
J.~Rapaport, Physics Letters B \textbf{92}, 233  (1980)

\bibitem{osterfeld_85}
F.~Osterfeld, V.A. Madsen, Phys. Rev. C \textbf{32}, 108 (1985)

\end{thebibliography}

\end{document}